\documentclass{aa}
\usepackage{graphicx}
\usepackage{txfonts}

\usepackage{natbib}         % this is the key to change the reference style (for a numbering format)

\usepackage{float}

\usepackage{capt-of}

\usepackage{stfloats}

\usepackage[switch]{lineno}           % line numbering for each column
\usepackage{soul}          % for Strikethrough
\usepackage{cancel}          % for Strikethrough
\usepackage{ulem}

\usepackage{hyperref}
\hypersetup{colorlinks,linkcolor={blue},citecolor={blue},urlcolor={blue}} 
\makeatletter
\renewcommand*\aa@pageof{, page \thepage{} of \pageref*{LastPage}}
\makeatother

\newcommand{\rv}[1]{{\color{black} #1}}
\newcommand{\rva}[1]{{\color{black} #1}}         % 09.06.2024
\newcommand{\rvc}[1]{{\color{black} #1}}      % 18.10.2024 -- cleaned version

\newcommand{\rvd}[1]{{\color{black} #1}}      % 05.11.2024 -- cleaned version

\newcommand{\rve}[1]{{\color{black} #1}}      % 11.11.2024 -- cleaned version

\newcommand{\rvf}[1]{{\color{black} #1}}      % 11.11.2024 -- cleaned version

\newcommand{\er}[1]{{\color{black} #1}}

\begin{document}

%\title{PKS\,1424-418: a persistent source of the radio--$\gamma$-ray connection?}
\title{\rv{PKS\,1424-418: A persistent candidate source of the \rvc{mm}--$\gamma$-ray connection?}}
%\title{The jet of PKS\,1424-418 as a persistent source of the radio--$\gamma$-ray connection?}

\titlerunning{
A long-term \rvc{mm}--$\gamma$-ray correlation in PKS\,1424-418
}

% collaborators
\author{Dae-Won Kim\inst{1},
Eduardo Ros\inst{1},
Matthias Kadler\inst{2},
Thomas P.\ Krichbaum\inst{1},
Guang-Yao Zhao\inst{1},
Florian R\"{o}sch\inst{2,1},
\newline
Andrei P.\ Lobanov\inst{1},
and J.\ Anton Zensus\inst{1}
% et al.\inst{2,3},
}

\authorrunning{D. -W. Kim et al.}

% 1
%---------------------
\institute{Max-Planck-Institut f\"{u}r Radioastronomie, Auf dem H\"{u}gel 69, 53121 Bonn, Germany\\ \email{dwkim@mpifr-bonn.mpg.de}
\and
% 2
%---------------------
%Institut f\"{u}r Theoretische Physik und Astrophysik, Universit\"{a}t W\"{u}rzburg, Emil-Fischer-Stra\ss{}e 31, 97074 W\"{u}rzburg, Germany
Julius-Maximilians-Universit\"{a}t W\"{u}rzburg, Fakult\"{a}t f\"{u}r Physik und Astronomie, Institut f\"{u}r Theoretische Physik und Astrophysik, Lehrstuhl f\"{u}r Astronomie, Emil-Fischer-Str. 31, D-97074 W\"{u}rzburg, Germany
%
%\and
% 3
%---------------------
%......TBD
}

%\date{Received September 15, 1800; accepted March 16, 1800}
\date{\today}

% \abstract{}{}{}{}{} 
% 5 {} token are mandatory
 
\abstract
% Note that the use of structured abstracts in A&A articles and Letters is not mandatory. Authors who prefer the traditional form are invited to implicitly follow the logical structure indicated above.
% context - aim - method - results - conclusion
{
We present a long-term strong correlation between millimeter (mm) radio and $\gamma$-ray emission in the flat-spectrum radio quasar (FSRQ) PKS\,1424$-$418. 
The \rvc{mm}--$\gamma$-ray connection in blazars is generally thought to \er{originate} from the relativistic jet close to the central engine.
% (i.e., supermassive black hole; SMBH).
%
We confirm a unique long-lasting \rvc{mm}--$\gamma$-ray correlation of \object{PKS\,1424$-$418} by using detailed correlation analyses and statistical tests, and we find its physical meaning in the source.
%
% We employed $\sim$8.5\,yr of radio and $\gamma$-ray light curves observed by the Atacama Large Millimeter/submillimeter Array; ALMA (Bands 3, 6, and 7; 90--350\,GHz) and the \textsl{Fermi}-large area telescope (LAT), respectively.
We employed $\sim$8.5\,yr of (sub)\rvc{mm} and $\gamma$-ray light curves observed by ALMA and \textsl{Fermi}-LAT, respectively.
From linear and cross-correlation analyses between the light curves, we found a significant, strong \rvc{mm}--$\gamma$-ray correlation over the whole period.
% At band 3 (90--100\,GHz), there is a very small delay of $\sim$3\,days (with 1\,$\sigma$ uncertainty of about 2\,days), which is $\gamma$-ray leading. At band 7 (345\,GHz), however, we do not find a notable time delay within uncertainty, thus meaning a zero-lag.
\rva{We did not find any notable time delay within the uncertainties for the \rvc{mm}--$\gamma$-ray correlation, which means zero lag.}
% within the data sampling intervals (i.e., 6\,days)
\rvc{The mm wave} spectral \rvc{index values (S$_{\nu}\,\propto\,\nu^{\alpha}$)} between the band 3 and 7 flux densities indicate a time-variable opacity of the source at (sub)mm wavelengths. Interestingly, the \rvc{mm wave} spectral index becomes temporarily flatter (i.e., $\alpha$\,$>$\,$-$0.5) when the source flares in the $\gamma$-rays.
% S$_{\nu}\,\propto\,\nu^{\alpha}$
We \er{relate} our \er{results} with the jet of \object{PKS\,1424$-$418}, and we discuss the origin of the $\gamma$-rays and opacity of the inner (sub)parsec-scale jet regions.
% \rev{CONCLUSION?}
%
}


\maketitle
%
%-------------------------------------------------------------------

\section{Introduction}
\label{intro}
% blazar nonthermal emission
%__________________________
Blazars are a subclass of radio-loud active galactic nuclei (AGNs). 
%
%Based on their optical spectra (i.e., with or without prominent emission lines), it has been divided further into flat-spectrum radio quasar (FSRQ) and BL\,Lac object, respectively 
\rv{The vast majority of bright $\gamma$-ray sources detected by the \textsl{Fermi}-Large Area Telescope (LAT) are identified as AGNs \citep[e.g., $\geq$\,80\% reported in][]{ajello2020}, and blazars dominate the $\gamma$-ray bright AGN population.}
Blazars can be further categorized into two subtypes: flat-spectrum radio quasars (FSRQs), and objects of BL Lacertae type (BLLacs). FSRQs show prominent optical emission lines, whereas BLLacs are without this feature 
\citep[see also][for a recent study of the dichotomy on weak and strong jets]{keenan2021}. 
Blazars are strong nonthermal sources, and the typical spectral energy distribution (SED) of blazars consists of two hump-like features,
%that one indicates the synchrotron process at radio-to-UV and the other one is the inverse-Compton process (IC) at $\gamma$-rays;
one of which is indicative of the synchrotron process from the radio to the UV, and the other is often presumed to be the inverse-Compton (IC) process in the $\gamma$-ray regime.
% \citep[see also][for hadronic models on the high-energy part]{bottcher2007}. 
\rv{Alternatively, hadronic models have also been suggested for the $\gamma$-ray emission \citep[e.g.,][]{dar1997, bottcher2007, bottcher2013}.}
X-ray bands can show a concave spectrum that is produced by both radiative processes \citep[e.g.,][]{magic2018}.
%
% The vast majority of bright $\gamma$-ray sources in the sky are identified as AGNs \citep[e.g., $\geq$\,80\% reported in][]{ajello2020} and blazars represent most of the sources in the $\gamma$-ray bright AGN population. 
%
The relativistic jets in blazars are considered to be the main source of the nonthermal emissions, including in the $\gamma$-ray band \citep[see e.g.,][for a physical picture of the process]{marscher2008}. 
Due to the small viewing angles between the jet axis and the line of sight
(e.g., typically $\leq$\,5$^{\circ}$),
%\rva{(e.g., around 5$^{\circ}$)},
emission from blazar jets is highly Doppler boosted and shows strong variability with time. An obvious question here is where and how the $\gamma$-rays are produced in the jets \citep{blandford2019}.

% Jet
%__________________________
%With the development of very-long baseline interferometry (VLBI), it has been revealed the detailed morphologies, structures, and physical properties of the (sub)parsec-scale jets of blazars 
With the development of very long-baseline interferometry (VLBI), detailed morphologies, structures, and physical properties of the (sub)parsec-scale jets of blazars were revealed 
\citep[e.g.,][]{lee2008, lister2009, nair2019, weaver2022}. 
In most of the VLBI images, the jets show a compact intense emitting region at the upstream end of the flow. This structure is called the radio core (or VLBI core), which is generally assumed to be the surface of the unity optical depth: $\tau_{\rm opaq}(\nu)\,\sim\,1$. This leads to the so-called core shift in the jets and tends to decrease with increasing frequency in a power-law form \citep[see e.g.,][]{dodson2017}. 
Hence, the innermost jet regions closer to the central engine are rather hidden and invisible owing to synchrotron self-absorption (SSA), particularly at lower radio frequencies (e.g., \citealt{lobanov1998, marscher2008}; see also \citealt{marscher2016} for alternative theories and models of the core). 
It has been suggested that the core shift is variable with time and can be increased by strong flares at higher radio frequencies \citep[e.g.,][]{chamani2023}.
In this context, observations at higher radio frequencies are crucial to broaden our understanding of the high-energy emission processes in blazar jets.
%
% Previous studies of the radio--$\gamma$-ray connection suggested that the $\gamma$-ray events are correlated with the radio core
Previous studies of the jets found that the $\gamma$-ray emission is correlated with the core radio emission
%\citep[e.g.,][]{jorstad2001a, kovalev2009, pushkarev2010}
\citep[e.g.,][\rvc{but see also \citealt{cheung2007, hodgson2018}, for the $\gamma$-rays with a different origin in the jets of some radio galaxies}]{jorstad2001a, kovalev2009, pushkarev2010}.
Based on VLBI observations, it was suggested that the $\gamma$-ray events are tightly linked with the passage of superluminal knots through the radio core region \citep[e.g.,][]{jorstad2001b, marscher2008, wehrle2016, roder2024}.

% time-correlation
%__________________________
\er{The search} for \er{time correlations} between radio and $\gamma$-ray light curves is one of the key approaches to support the \er{existence} of the radio--$\gamma$-ray connection in blazars. Previous studies performed a statistical study with a large number of AGN samples (mostly blazars) and found that just a handful of sources show a significant radio--$\gamma$-ray correlation over a short period (i.e., 2.5--5.0\,yr) of time \citep[][]{fuhrmann2014, maxm2014a, rama2015, rama2016}. \er{In contrast to} the clear correlation between the optical and $\gamma$-ray emission in blazars \citep[e.g.,][]{liodakis2019}, it is difficult to find a clear significant correlation between radio and $\gamma$-ray emission.
% ; \er{especiallly}, for sources with high variability \er{in} both wavebands \citep[e.g.,][]{kim2022}.
This might be due to (1) different variability timescales, (2) a lack of huge flares, and (3) poor data sampling.
% The typical radio--$\gamma$-ray correlations can mainly be distinguished by either $\gamma$-ray leading \citep[e.g.,][]{agudo2011a, kim2020} or (quasi-)simultaneous \citep[e.g.,][]{wehrle2012, kim2018}.
\rvf{The typical radio--$\gamma$-ray correlation is either (1) the $\gamma$-rays preceding the radio emission \citep[e.g.,][]{agudo2011a, kim2020} or (2) the $\gamma$-rays (quasi-)simultaneous with the radio emission \citep[e.g.,][]{wehrle2012, kim2018}.
}
% without a long time delay\LEt{***this sentence is ungrammatical as it reads now and I don't quite know how to fix it for you. Please check this again and rephrase for grammatical correctness and clarity***}
This suggests that the $\gamma$-ray production site is located either in a region upstream of or near the radio core \citep[but see also][for a scenario of a region downstream from the core]{leon2011}. In addition,
% \LEt{***your use of the slash is ambiguous here: do you mean "and" or "or"? Please use words for clarity when the slash doesn't mean "ratio", as in "S/N" for "signal-to-noise ratio", for instrument pairings, and for "and/or". Please check this throughout and change as required. I'll not highlight this again to avoid cluttering the ms***}
previous \rvf{and} recent works reported multiple sites for $\gamma$-rays in the jets \citep[e.g.,][]{marscher2008, rani2018, liodakis2020, kim2022}.

% PKS1424-418
\object{PKS\,1424$-$418} \citep[$z$\,$\sim$\,1.52;][]{white1988} is an FSRQ in the southern sky (DEC\,$\sim$\,$-$42$^{\circ}$). Since most of the major VLBI arrays are located in the northern hemisphere, there are only a limited number of VLBI images of this source. The jet of \object{PKS\,1424$-$418} has an extended structure toward the northeast in the image plane that indicates the flow axis \citep[see][]{benke2024}.
%
% \object{PKS\,1424$-$418} is bright at both radio and $\gamma$-ray bands, and as expected, the compact radio core is the brightest feature in the jet.
%
Interestingly, a PeV-energy neutrino event was detected from this source
along with strong $\gamma$-ray outbursts in late 2012 to early 2013 \citep{kadler2016}. This implies the presence of hadronic processes (e.g., photo-pion production with highly accelerated protons) in the jet of \object{PKS\,1424$-$418}.
% \citet{kadler2016} reported a PeV-energy neutrino event along with strong $\gamma$-ray outbursts detected from the source in late-2012 to early-2013.
% Hadronic processes (e.g., photo-pion production in the presence of highly accelerated protons) can be responsible for this neutrino event.
% In the meantime, some of previous works on \object{PKS\,1424$-$418} suggested leptonic models as the dominant radiative processes in the jet by using multi-waveband (i.e., IR, optical, and $\gamma$-rays) light curve data \citep[e.g.,][]{buson2014, abhir2021}.

\rvc{In this work, we define the spectral index ($\alpha$) as S$_{\nu}\,\propto\,\nu^{\alpha}$, where S and $\nu$ are the flux density and observing frequency, respectively.
\rvd{We estimated the uncertainty of $\alpha$ following the error propagation.}
We use the cosmological parameters as follows: $H_{0}$\,=\,71\,km\,Mpc\,s$^{-1}$, $\Omega_{\Lambda}$\,=\,0.73, and $\Omega_{m}$\,=\,0.27 throughout this paper.
}

%============================================================================================
%============================================================================================
%============================================================================================

\section{Radio and \texorpdfstring{$\gamma$}{g}-ray light curves in 2011--2020}
%\section{\rvc{The light curves} in 2011--2020}
\label{sec:lcdata}
%
%__________________________
% LAT
%\section{\textsl{Fermi}-LAT data calibration}
%\section{\textsl{Fermi}-LAT data \rva{reduction}}
%\label{sec:gcalib}

\subsection{\rvc{\texorpdfstring{$\gamma$}{g}-ray data}}
\label{sec:gamdat}
We analyzed the Pass 8 $\gamma$-ray data obtained from the \textsl{Fermi}-LAT \citep{atwood2009}. 
Overall, we followed the standard unbinned likelihood procedure\footnote{\url{https://fermi.gsfc.nasa.gov/ssc/data/analysis/scitools/}} to generate the $\gamma$-ray light curve of the source (i.e., RA\,=\,216.987$^{\circ}$, DEC\,=\,$-$42.106$^{\circ}$; J2000). 
%
% some templates/catalog + background modelling (xml)
At first, we put together a background template (so-called XML model). Two background components, \texttt{gll\_iem\_v07} and \texttt{iso\_P8R3\_SOURCE\_V3\_v1}, were employed to take into account Galactic diffuse emission and isotropic background emission, respectively. 
We set a region of interest (ROI) to be a circle of radius 10$^{\circ}$ centered at the position of \object{PKS\,1424$-$418}, and we included all
% \LEt{***please provide the spelled-out names of all instruments and surveys at first occurrence in the main text. This may be done as a footnote if it interrupts the flow of the sentence too much. Please check and amend as required***}
4FGL-catalog sources \citep[4FGL-DR4,][]{abdollahi2020, ballet2023} present within 20$^{\circ}$ (ROI\,$+$\,10$^{\circ}$) from the position of \object{PKS\,1424$-$418}. We kept the default spectral models provided by the catalog for all the sources except our target; a simple power law was applied to the target (see below).

During the spectral fits, we left the spectral parameters free for all sources within 5$^{\circ}$ of the ROI center. For sources between 5$^{\circ}$ and 10$^{\circ}$, only normalization parameters were left free. For sources beyond 10$^{\circ}$, we kept all the parameters fixed to the catalog values. For the two diffuse background components (i.e., Galactic and isotropic), only the normalizations were left free. When sources lay within 10$^{\circ}$ but were faint (i.e., $<$\,5\,$\sigma$, where $\sigma$ is the detection significance), then their spectral parameters were kept fixed; one source was known to be time variable (i.e., variability index $>$\,24.73\rvc{; see, e.g., \citealt{abdollahi2023} for a definition of the variability index}), and thus, we only let its normalization be free.

\begin{figure}[t]
\centering
\includegraphics[angle=0, width=\columnwidth, keepaspectratio]{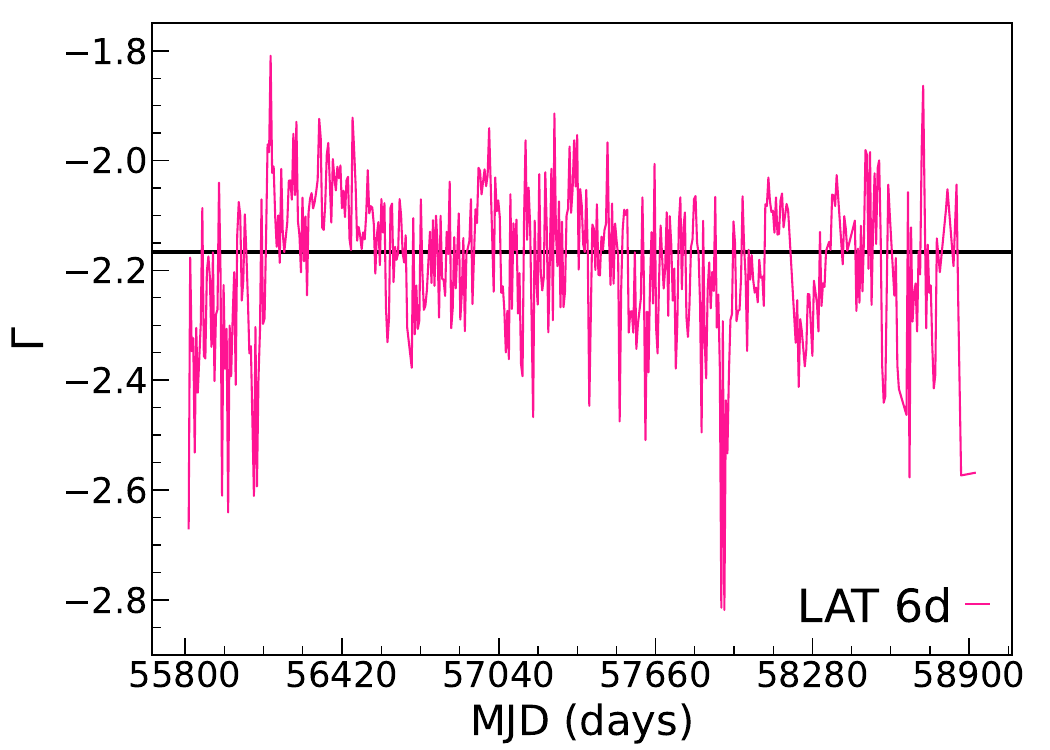} \
\includegraphics[angle=0, width=\columnwidth, keepaspectratio]{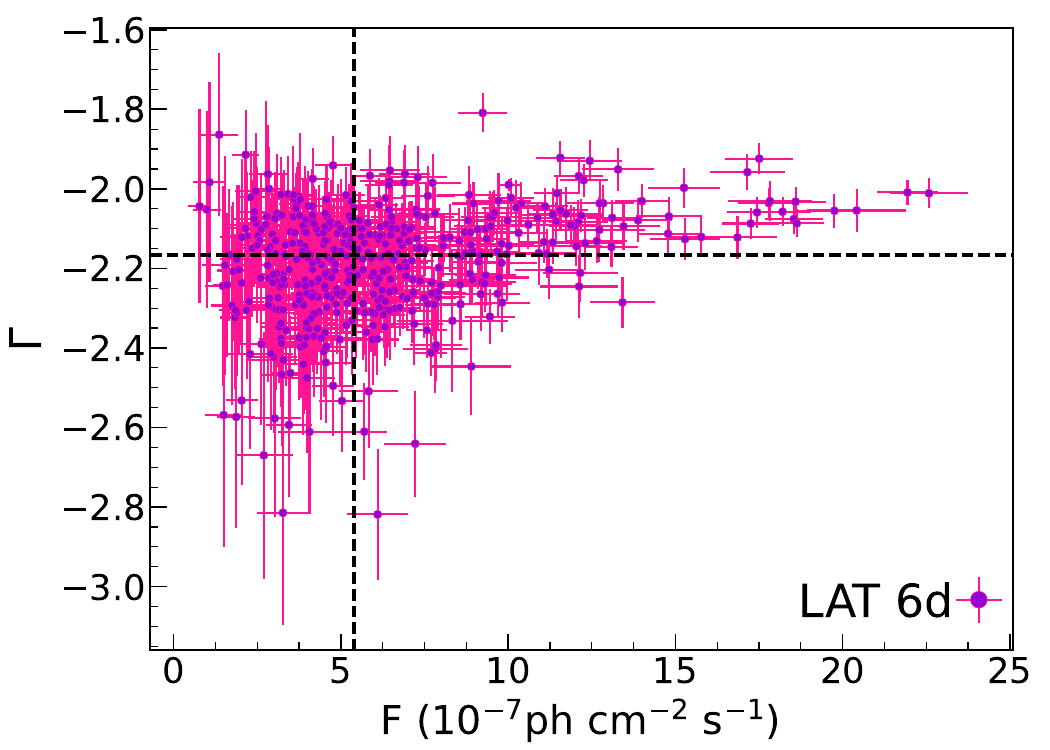}
\caption{
Power-law index vs. time (\textsl{upper}) and flux density (\textsl{lower}) of the $\gamma$-ray emission of \object{PKS\,1424$-$418}. The black lines in each panel indicate the median values on each axis: $\Gamma_{\rm median}$\,=\,$-$2.17 and F$_{\rm median}$\,=\,$5.39\times10^{-7}$\,ph\,cm$^{-2}$\,s$^{-1}$.
}
\label{fig:gindx}
\end{figure}

% overall procedure
We analyzed $\gamma$-rays from the source in the 0.1--200\,GeV energy range measured between 4 September 2011 and 26 March 2020 (MJD: 55808--58934), which is $\sim$8.5\,yr.
\rvc{We chose this time range because (1) the ALMA light curves of this source started in 2011 and (2) the source became quiescent in the LAT $\gamma$-ray band for a long time (i.e., $\sim$1.5\,yr) with too many upper limits from 2020.
}
We selected SOURCE class (\texttt{evclass=128}) and FRONT$+$BACK-type (\texttt{evtype=3}) events.
The good-time intervals (GTIs) were determined with the parameters \texttt{(DATA\_QUAL>0)\&\&(LAT\_CONFIG==1)} and \texttt{roicut=no}. The maximum zenith angle (zmax) was set to be 90$^{\circ}$ to minimize the Earth limb $\gamma$-ray contamination.
% More technical details about the LAT data selection can be found in the webpage\footnote{\url{https://fermi.gsfc.nasa.gov/ssc/data/analysis/documentation/Cicerone/Cicerone_Data_Exploration/Data_preparation.html}}.
We employed the maximum likelihood test statistics (TS) to evaluate the significance of the $\gamma$-ray photons. For each time bin, we first fit the data with the background template. Then, all the faint sources with TS\,$<$\,9 ($\sim$3\,$\sigma$) were removed from the background model. Using this updated model, we performed a second maximum likelihood fit to the data to obtain the final results. We computed 2\,$\sigma$ upper limits for the $\gamma$-rays with TS\,$<$\,20 or $\Delta F_{\gamma}/F_{\gamma}$\,$\geq$\,0.5, where $F_{\gamma}$ and $\Delta F_{\gamma}$ indicate the $\gamma$-ray flux density and its uncertainty, respectively.

We used a power-law form that was defined as $dN/dE \propto E^{\Gamma}$, with $N$ being the number of photons, $E$ the photon energy, and $\Gamma$ the photon index.
\rv{For the LAT analysis, the power-law and LogParabola models\footnote{\url{https://fermi.gsfc.nasa.gov/ssc/data/analysis/scitools/source\_models.html}} are generally preferable for AGNs. In our case, there are no \rvc{significant} differences in the results between the two models, and we selected the power-law model assuming that it is more suitable for these short day-scale bins.}
Since we let $\Gamma$ vary during the fits, every single time bin had a different $\Gamma$ value. Figure~\ref{fig:gindx} shows the results. Overall, $\Gamma$ fluctuates in the time domain without a clear tendency. However, $\Gamma$ seems to converge \er{to about} $-$2.0 with increasing flux density.
% The overall index--flux trend of \object{PKS\,1424$-$418} is quite similar to the case of \object{3C\,273} reported in \citet{kim2020}, but brighter and harder.
\rv{The overall index--flux trend of \object{PKS\,1424$-$418} is similar to a typical pattern for \rvc{$\gamma$-ray emitting FSRQs} \citep[see e.g.,][for the case of \object{3C\,273}]{kim2020}, but fairly harder and brighter.}
\rvd{To match the $\gamma$-ray data with the sampling of our radio data, we generated a $\gamma$-ray light curve with a binning interval of 6\,days (see Section~\ref{sec:raddat}).}

\begin{figure*}[t]%[!htbp] 
% just put this very earlier than its position supposed to be..
\centering
\includegraphics[angle=0, width=\textwidth, keepaspectratio]{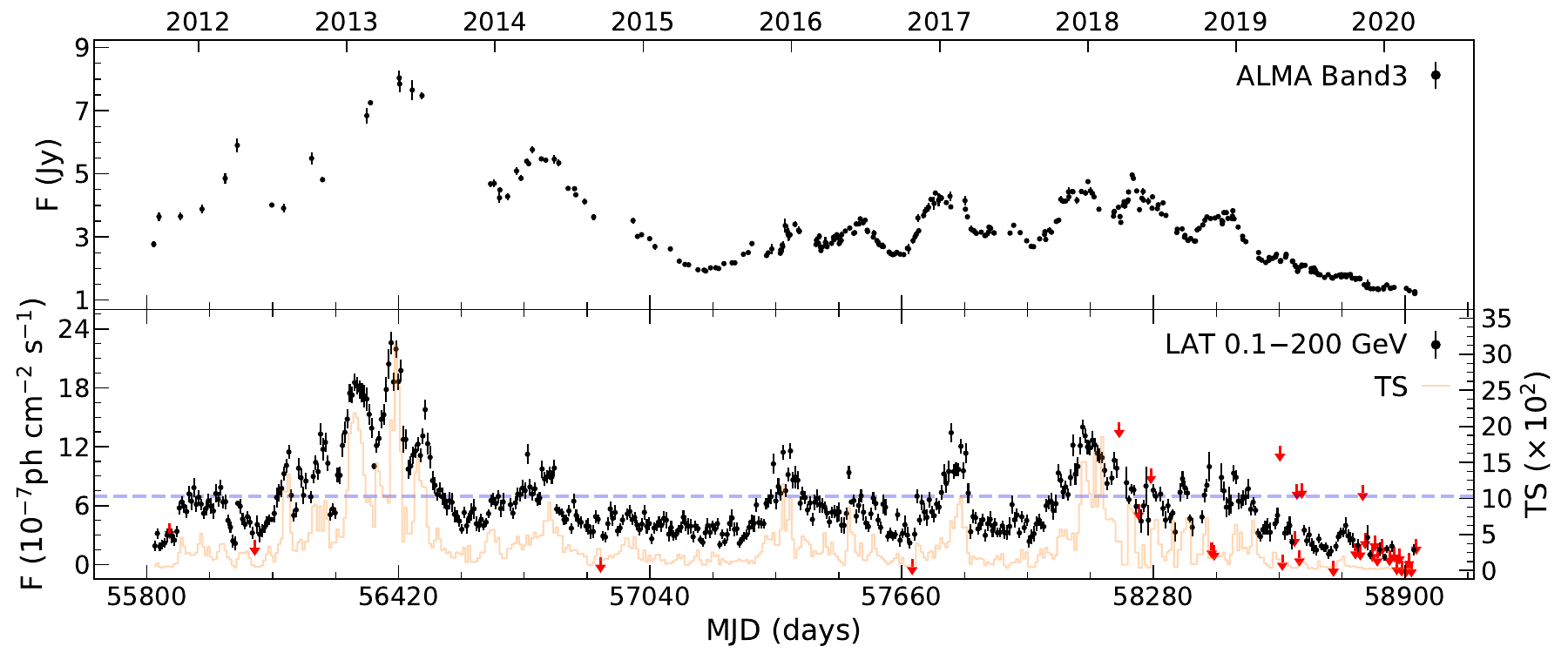}
\caption{
ALMA band 3 (95\,GHz) light curve (\textsl{upper}) and 6-day-binned LAT $\gamma$-ray light curve (\textsl{lower}) of \object{PKS\,1424$-$418} in 2011--2020 ($\sim$8.5\,yr). The red arrows indicate the 2\,$\sigma$ upper limits of the $\gamma$-rays. 
% Significance of the $\gamma$-ray photons (test statistics; TS) 
\rv{The detection of $\gamma$-ray photons in terms of TS}
is shown as the light orange line. \rvc{The horizontal blue line denotes the threshold for the $\gamma$-ray flare.}
}
\label{fig:thelcs}
\end{figure*}

In total, we obtained
522 $\gamma$-ray photon measurements, including 14 bad time bins
for which the fit was unsuccessful with some fitting errors. We excluded these poor data in this work. Only data points with TS\,$\geq$\,20 and $\Delta F_{\gamma}/F_{\gamma}$\,$<$\,0.5 were used in this work, which is $\sim$93\% (470/508) of the whole samples. For the other poor data points, we present the upper limits only in Figure~\ref{fig:thelcs}, and they were excluded in the rest of our results.
During the source analysis, we occasionally found that the fit returned erroneous results (e.g., empty or highly underestimated bins) depending on the fit-tolerance (Tol) value. To remedy this issue, we carried out two sets of the analysis: one set with Tol\,$\sim$\,0.01, and the other set with Tol\,$\sim$\,0.00001. When they both returned reasonable estimates, we chose the set with a higher TS or lower $\Delta F_{\gamma}/F_{\gamma}$ when the difference in TS was smaller than one.
\rvc{In our $\gamma$-ray light curve, we defined the low-state photons as those with TS\,$<$\,100 (76 photons in total). They are $\sim$3$\rm \,\times\,10^{-7}\,ph\,cm^{-2}\,s^{-1}$ on average, with 1\,$\sigma$ being $\sim$1.3$\rm \,\times\,10^{-7}\,ph\,cm^{-2}\,s^{-1}$. Based on this estimate, we considered all those photons brighter than $\sim$7$\rm \,\times\,10^{-7}\,ph\,cm^{-2}\,s^{-1}$ (i.e., $\geq$ the average~$+$~3\,$\sigma$) as a flare.
}
% 6.965386541132714e-07

\subsection{\rvc{Millimeter-wave data}}
\label{sec:raddat}
% ALMA
We made use of the Atacama Large Millimeter/submillimeter Array (ALMA) calibrator catalog data\footnote{\url{https://almascience.eso.org/alma-data/calibrator-catalogue}}. We obtained three ALMA \rvc{millimeter (mm)} radio light curves of \object{PKS\,1424$-$418} at band 3, band 6, and band 7 (see more details of the ALMA bands here\footnote{\url{https://www.eso.org/public/teles-instr/alma/receiver-bands/}}). At band 3, the flux densities were measured at 
% a frequency range of $\sim$91\,GHz to $\sim$103\,GHz.
\rv{two central frequencies: $\sim$91\,GHz and $\sim$103\,GHz.}
We used all these measurements and consider 95\,GHz as a representative observing frequency of band 3 in this study. When multiple flux density measurements were available at the same timestamp, we used their average. The other two bands (i.e., bands 6 and 7) observed the source at
% $\sim$235\,GHz and $\sim$345\,GHz
\rv{the central frequencies of $\sim$235\,GHz and $\sim$345\,GHz, respectively.}
Details of the catalog and ALMA observations can be found in \citet{bonato2018}.
%
% The LAT has been carrying out all-sky survey at $\gamma$-rays (i.e., from tens of MeV to $>$\,300\,GeV), since 2008 \citep{atwood2009}. It offers unique opportunities for exploring the nature of $\gamma$-ray bright sources, particularly blazars (see e.g., a LAT monitored source list\footnote{https://fermi.gsfc.nasa.gov/ssc/data/access/lat/msl\_lc/}). For data analysis, we fully calibrated raw LAT data of \object{PKS\,1424$-$418}. For details of our LAT data calibration, we refer to Appendix~\ref{sec:gcalib}. 

% light curves
In our analyses, the cadence \rv{(i.e., the sampling frequency)} is the most important factor for obtaining more accurate results, and we found that the band 3 light curve was best sampled: The median sampling interval is 6\,days, with an average of $\sim$10\,days (see Appendix~\ref{sec:allalma} for the band 6 and 7 data).
% To match the $\gamma$-rays with the sampling of the band 3 data, we generated a $\gamma$-ray light curve with a binning interval of 6\,days
For all the ALMA bands, the medians describe their samplings better than the average
\rv{(i.e., closer to the peak frequency in the sampling distribution).}
% Hence, we further focus mainly on analyzing a connection between the band 3 and $\gamma$-ray data in this study.
%\rv{Since the high cadence of the band 3 compared to the other two radio bands,
Since band 3 had a higher cadence than the other two radio bands,
our analyses were more focused on
the data from this band together with the LAT $\gamma$-ray data.
%
%We took $\sim$8.5\,yr of the ALMA and LAT data, starting from September 2011 to March 2020.
\rvd{As for the \textsl{Fermi}-LAT data analysis, we considered ALMA data spanning $\sim$8.5\,yr from September 2011 to March 2020.}
Figure~\ref{fig:thelcs} shows the band 3 and $\gamma$-ray light curves.
The band 3 flux densities are $\sim$3\,Jansky (Jy) on average at 95\,GHz with 1\,$\sigma$ of 1\,Jy.
% In the case of the LAT $\gamma$-ray data, the average photon flux is $\sim$6$\rm \,\times\,10^{-7}\,ph\,cm^{-2}\,s^{-1}$ at 0.1--200\,GeV with 1\,$\sigma$ being $\sim$4$\rm \,\times\,10^{-7}\,ph\,cm^{-2}\,s^{-1}$.
From simple visual inspection of the light curves, it seems already obvious that their overall patterns look very similar to each other, 
%thus meaning radio and $\gamma$-ray variability of the source.
which implies a connection between the \rvc{mm} and $\gamma$-ray variability. 
% the time range
% 337564802.0 338083202.0 3 55820.000766018515 6.0
% 605577602.0 606096002.0 520 58922.000766018515 6.0

\begin{figure}[t]
\centering
\includegraphics[angle=0, width=\columnwidth, keepaspectratio]{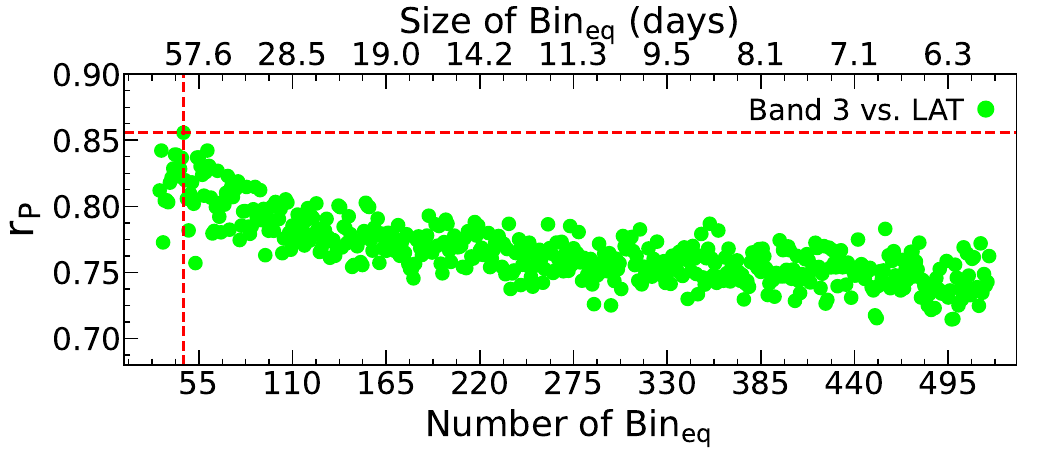} \
\includegraphics[angle=0, width=\columnwidth, keepaspectratio]{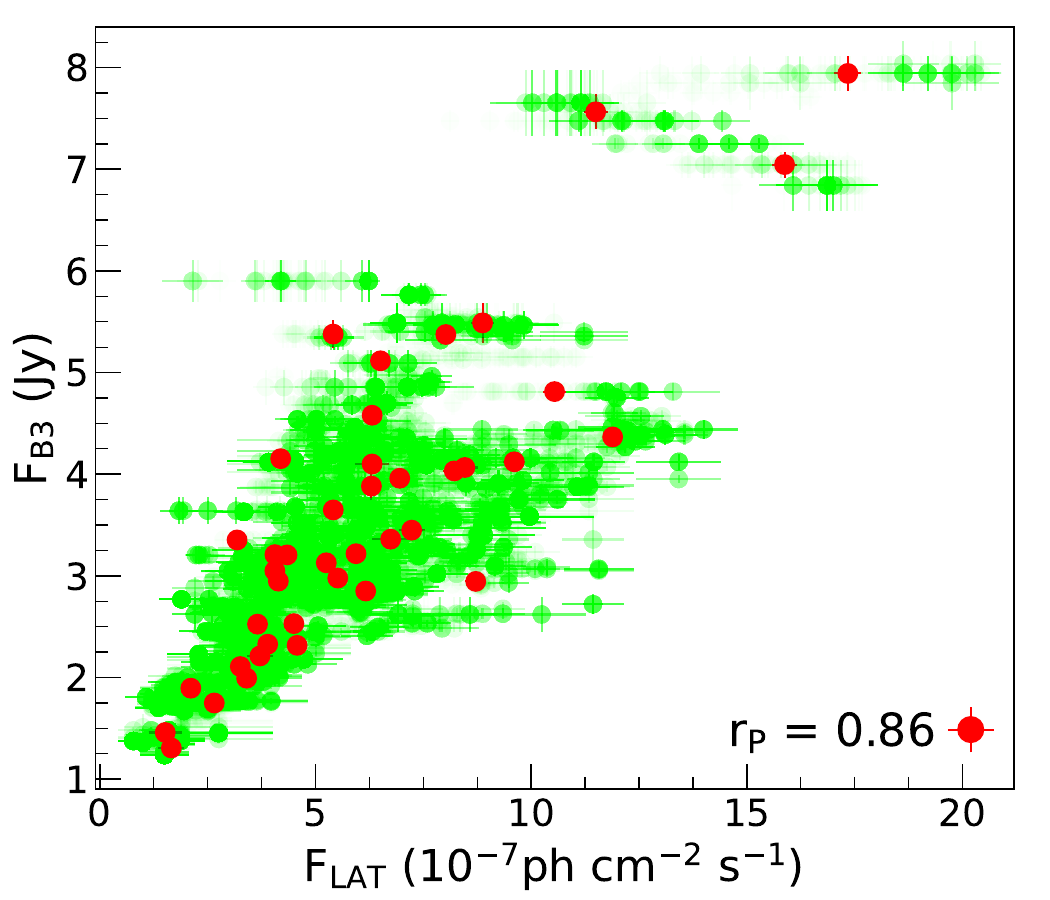}
\caption{
Pearson correlation tests between the ALMA band 3 and LAT light curves for the whole 8.5\,yr period. We estimated 488 different Pearson coefficients (r$_{\rm P}$) using different time-grids (see Appendix~\ref{sec:panal}). Upper panel:  r$_{\rm P}$ distribution on different scales of the time bin. The dashed red lines refer to the highest r$_{\rm P}$ value (i.e., $\sim$0.86) at Bin$_{\rm eq}$\,=\,69.1\,days (NBin$_{\rm eq}$\,=\,46). Lower panel: Flux--flux scatters corresponding to the r$_{\rm P}$ estimates shown in the upper panel. The scatter for the highest r$_{\rm P}$ is shown in red. The transparency of the data points is set to be very high (thus faint) to highlight the points that appear more frequently.
}
\label{fig:coreps}
\end{figure}
% ...with N=46 (corresponds to DT=69.06) --> overlapping 41/45 (91%), then the Pearson = 0.86!

%=======================================================================
\section{Results}
\label{sec:result}
% suggestion by GYZ
% Overall, the radio light curves were sampled unevenly and distributed randomly; in the case of the $\gamma$-ray data, there are some empty time bins (i.e., erroneous data) and upper limits, thus meaning that its sampling is not completely regular. 
\rv{While we estimated the median sampling times at the three ALMA bands, the radio light curves were sampled unevenly because the cadence varied, as shown in Figure~\ref{fig:thelcs}. The $\gamma$-ray data, on the other hand, have some empty time bins (i.e., erroneous data) and upper limits that lead to irregular sampling.}
%
% To reflect this issue in our analyses properly,
\rv{To properly account for this in our analyses,}
we developed a binning approach (see Appendix~\ref{sec:panal}), and this approach was used throughout this paper. A connection between the \rvc{mm} and $\gamma$-ray light curves was examined by using the Pearson correlation and the local cross-correlation function \citep[LCCF; e.g.,][]{welsh1999}. Detailed descriptions of these methods can be found in Appendix~\ref{sec:panal} and Appendix~\ref{sec:ccf1}, respectively. 
% Lin.
% to take this sampling issue into account, we came up with a binning approach to calculate Pearson correlation coefficient (rP) between the band 3 and LAT fluxes (see Appendix C, for details of the method).
% Crs.
% We employed the Local Cross-Correlation Function (LCCF; see Appendix D, for detailed description of the method and a comparison with the DCF function) to further investigate the \rvc{mm}–γ-ray correlation

\subsection{Linear relations}
\label{sec:pear}
The Pearson correlation is a straightforward way to examine the presence of any significant linear correlation between two time series. 
Figure~\ref{fig:coreps} shows the results. We obtained a total of 488
\rv{Pearson coefficients (r$_{\rm P}$)}
for the whole 8.5\,yr of the \rvc{mm} and $\gamma$-ray variability. Bin$_{\rm eq}$ (i.e., the equal-sized time bin) varies with the 488 calculations from 6\,days to 100\,days. The r$_{\rm P}$ values are $\sim$0.76 on average, with a 1\,$\sigma$ \rv{uncertainty} of 0.02 (minimum and maximum of 0.71 and 0.86, respectively). A sufficient number of data samples was used for each of the calculations, and thus, all the r$_{\rm P}$ values are highly significant
% (i.e., $p$-value: $>$\,99.9\%).
\rv{(i.e., $p$-values $\ll$ 0.01)}.
% 32 to 519
%
As expected from the visual inspection of Figure~\ref{fig:thelcs}, there is a clear and strong positive correlation between the band 3 and $\gamma$-ray emission in \object{PKS\,1424$-$418} in all observations. 
A level of r$_{\rm P}$ = 0.76 is already quite high; even the minimum value of 0.71 is also high enough. 
Overall, the impact of different Bin$_{\rm eq}$ sizes on r$_{\rm P}$ is shallow, but it is stronger above $\sim$30\,days. 
%The impact of different Bins$_{\rm eq}$ sizes on r$_{\rm P}$ is not significant below $\sim$30\,days. 
%seems rather weak at least below, for instance, $\sim$30\,days. 
%
This strong significant linear relation in the presence of multiple flaring activities at both wavebands for a long time indicates that the mm wave and $\gamma$-ray emissions are physically connected with each other \citep[e.g.,][]{wehrle2012}.
In Appendix~\ref{sec:allalma}, we also test Pearson correlations between the three radio bands (i.e., band 3, 6, and 7). As a result, we find that they are tightly correlated almost one to one with r$_{\rm P}$ of $>$\,0.95.
% radio relationships in the appendix

\begin{figure}[b]
\centering
\includegraphics[angle=0, width=\columnwidth, keepaspectratio]{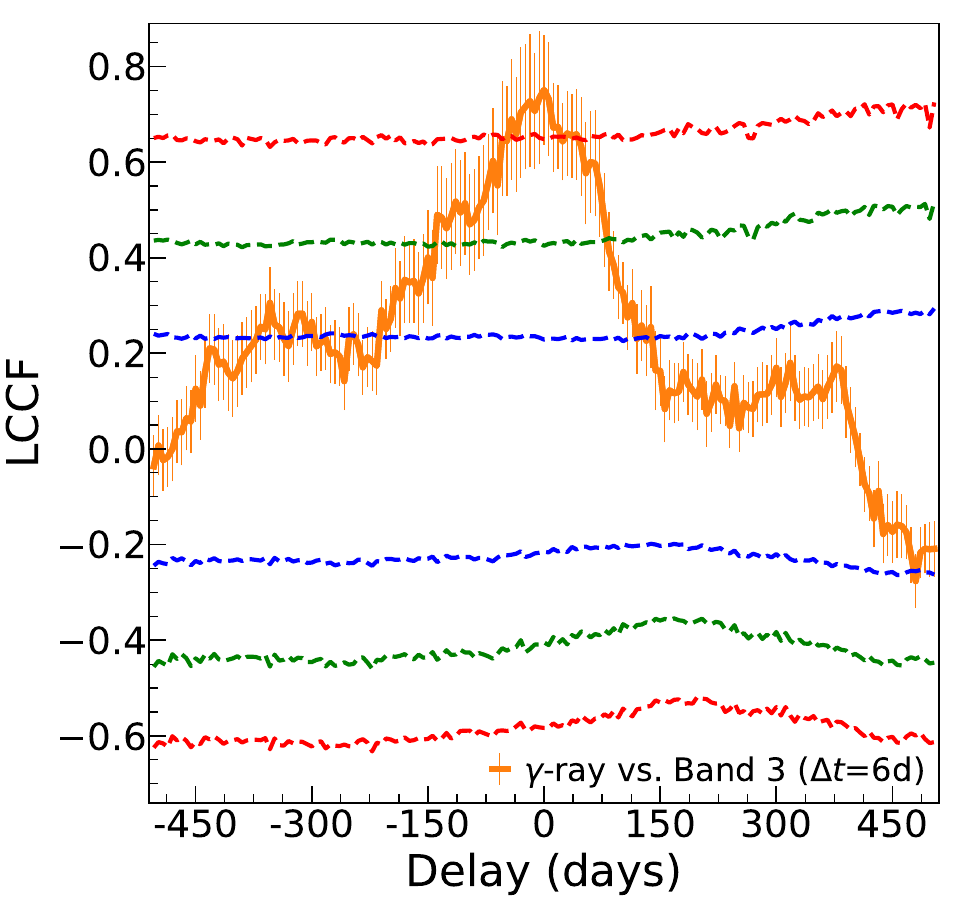}
\caption{
Cross-correlation (LCCF) curve between the band 3 and $\gamma$-ray light curves for the whole 8.5\,yr period. The dashed blue, green, and red lines are the 68\%, 95\%, and 99.9\% confidence levels, respectively.
}
\label{fig:lccfmain}
\end{figure}

\subsection{Cross-correlation between \rvc{mm} and \texorpdfstring{$\gamma$}{g}-ray emission}
\label{sec:cros}
The linear correlations we presented in the previous section were obtained without any time shift of a light curve (either \rvc{mm} or $\gamma$-ray). This \er{ suggests that there is} no significant time delay in the physical connection between the mm wave and $\gamma$-ray emission in the source. However, the delay plays an important role in the physical interpretation of the results, and \er{so} we carried out a detailed \er{search} for it by using LCCF. 
% We employed the Local Cross-Correlation Function (LCCF; see Appendix~\ref{sec:ccf1}, for detailed description of the method and a comparison with the DCF function) to further investigate the \rvc{mm}--$\gamma$-ray correlation. 
% (see also Appendix~\ref{sec:ccf1}, for a comparison with the DCF function) to further investigate the \rvc{mm}--$\gamma$-ray correlation.
Figure~\ref{fig:lccfmain} shows the resultant LCCF curve between the band 3 and $\gamma$-ray light curves of \object{PKS\,1424$-$418}. The binning interval ($\Delta t$) was set to be 6\,days to match the basic samplings of the band 3 and $\gamma$-ray data; we also tested higher $\Delta t$ values, but the results were consistent with the one with $\Delta t$ = 6\,days (see Appendix~\ref{sec:ccf2}). The LCCF values are mostly above zero for all delays ($\tau$), and we found one notable hump-like feature at around $\tau$\,=\,0. Only the top part of this central LCCF hump where it peaks at $\tau$ = 0 with LCCF\,$\sim$\,0.75, exceeds the confidence level of 99.9\%.
%To better constrain the peak position (i.e., $\tau$ of the center of the significant LCCF hump), we fit a single Gaussian to it. We refer to Appendix~\ref{sec:ccf2} for details of the fitting process (also for the LCCF curve between the $\gamma$-ray and band 7 data). We find that the Gaussian peaks are approximately at $-$2.8$\pm$2.0\,days (i.e., ``$\gamma$-ray leading'') for the band 3 hump and $-$0.8$\pm$2.6\,days for the band 7 hump.
%This suggests that there is a very small amount of the core shift between band 3 and 7 that could be present at such high radio frequencies \citep[see e.g.,][for a typical tendency of the core shift profiles]{dodson2017, lisakov2017, okino2022}. Given the estimated delays between the three bands (i.e., $\gamma$-rays vs. the bands 3 \& 7), we find that the $\gamma$-rays are simultaneous with the band 7 flux densities, but (quasi-)simultaneous with the band 3 flux densities; for band 3, a further investigation of radio spectral indices is needed and we deal with this issue in the next section.
\rva{This result suggests that the $\gamma$-ray and band 3 light curves are contemporaneous in the time domain with an uncertainty of $\pm$3\,days from $\tau$\,=\,0.}

\begin{figure*}[t]
\centering
\includegraphics[angle=0, width=\textwidth, keepaspectratio]{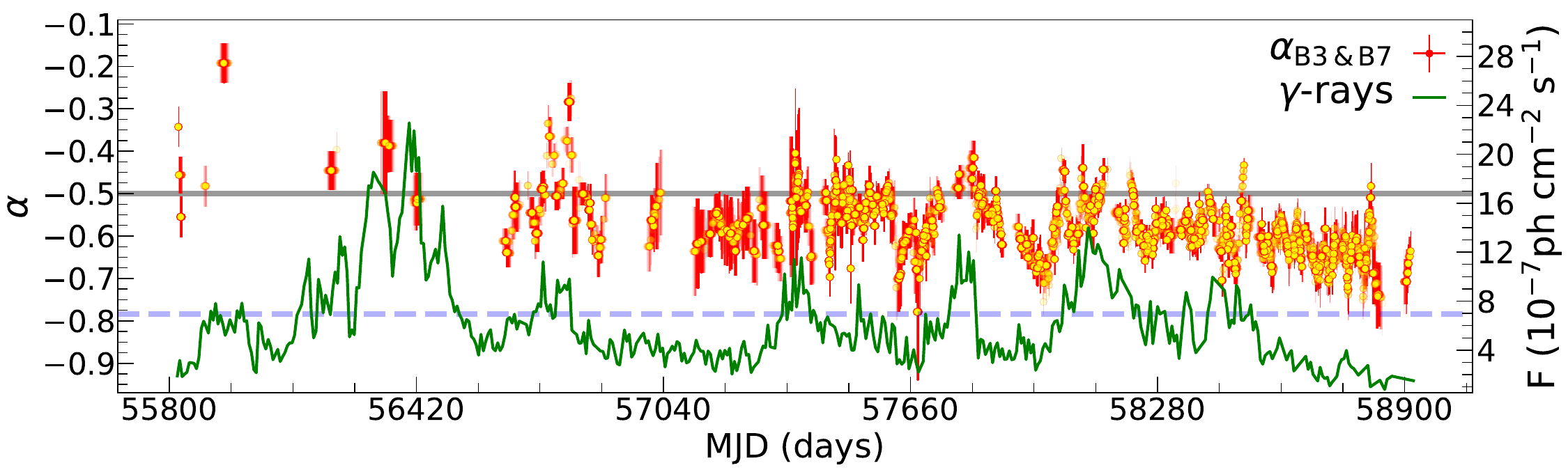}
\caption{
Evolution of the (sub)mm wave radio spectral index \rvc{($\alpha$)} between ALMA band 3 and 7 (i.e., 95\,GHz vs. 345\,GHz). The horizontal gray line points to $\alpha$ = $-$0.5. The $\gamma$-ray light curve is overlaid in the background for comparison. \rvc{The horizontal blue line denotes the threshold for the $\gamma$-ray flare.}
}
\label{fig:mmspidx}
\end{figure*}

\subsection{Evolution of the (sub)mm wave spectral index}
\label{sec:rspix}
%
% From our LCCF analysis (see Section~\ref{sec:cros}), we find a time delay of $\sim$3\,days between the band 3 and $\gamma$-ray data, whereas \rv{near-zero} delay with the band 7 data within uncertainty. These results are consistent with a typical systematic behavior of the \rvc{mm}--$\gamma$-ray correlation in blazars: smaller delays at higher radio frequencies \citep[e.g.,][]{fuhrmann2014}. The amount of $\sim$3\,days is actually a very small delay compared to previous results at lower radio frequencies in other sources (e.g., \citealt{leon2011, maxm2014a}; see also \citealt{paraschos2023}, \rv{for much larger delays of $\sim$1.5\,yr} even at $\geq$\,230\,GHz in \object{3C\,84}).
% 
% Fortunately, the band 7 data \er{were} quite well sampled (i.e., a median of $\sim$8\,days); \er{in contrast}, the sampling of the band 6 data is very poor which \er{prevents} us from making a \er{solid} analysis with it. 

\rva{Since the band 7 data were quite well sampled (i.e., a median of $\sim$8\,days), we also calculated the spectral index \rvc{($\alpha$)} between the band 3 and 7 flux densities to check the source opacity at (sub)mm wavelengths}; in contrast, the sampling of the band 6 data is very poor, which prevented us from using it for a \er{solid} analysis. 
%To check the source opacity at (sub)mm-wavelengths, we \er{also} calculated \er{the} spectral index ($\alpha$) between the band 3 and 7 fluxes (S$_{\nu}\,\propto\,\nu^{\alpha}$). 
Figure~\ref{fig:mmspidx} shows the evolution of $\alpha$ between the 95\,GHz and 345\,GHz emission.
The variability timescales of the radio light curves seem to be on month scales overall. To take this into account in the calculations of $\alpha$, the radio data were binned with multiple bin sizes in the time domain. That is, Bin$_{\rm eq}$ ranged from 3\,days to 30\,days, which corresponds to NBin$_{\rm eq}$ sliding from 1038 to 105.
\rvd{We also tested very small Bins$_{\rm eq}$ that varied from from one day to two days. The overall trend was consistent with the result from the larger Bin$_{\rm eq}$ range, but with far fewer numbers of the data points.}
%
% In the calculations, we set Bin$_{\rm eq}$ to be 3\,days to 30\,days which corresponds to NBin$_{\rm eq}$ sliding from 1038 to 105.
The average of the $\alpha$ values is $-$0.57 (with 1\,$\sigma$ of 0.08), and they range from a minimum of\ $-$0.78 to a maximum of $-$0.19.

We considered $\alpha$\,$\sim$\,$-$0.5 as a reference point of the transition between optically thick (flatter; $\alpha$\,$>$\,$-$0.5) and thin (steeper; $\alpha$\,$\leq$\,$-$0.5) \citep{marscher1985}. Most of the time, the spectral indices are below $-$0.5 at around $-$0.6.
\rvd{From visual inspection of Figure~\ref{fig:mmspidx}, however, we found that $\alpha$ becomes flatter (i.e., $\geq$\,$-$0.5) when the $\gamma$-rays flare.}
However, this feature seems to be clearer when the $\gamma$-ray flares are stronger and \rvd{last longer}. For the weaker flares, the spectral feature might be less pronounced in the thin regime (i.e., $\alpha$\,$<$\,$-$0.5), or they might have a different origin/mechanism without this feature in the (sub)mm wave spectrum.
\rvd{In early 2012 (around MJD\,55936), higher $\gamma$-rays were observed that perhaps consisted of multiple $\gamma$-ray flares. This activity seems marginal or exceeds the flare threshold only slightly. We consider it to be a long-lasting activity (e.g., $\geq$\,4\,months), but weaker.
}
% In [85]: 55936
% In [86]: '2012-01-10'
%
It is worth noting that there are a few caveats for Figure~\ref{fig:mmspidx}. We note that the sampling of the radio data at both bands 3 and 7 are irregular, and the sampling is quite sparse especially in the early part (i.e., $<$\,2014; MJD\,56658). Owing to this, the $\alpha$ values in this early part might be less robust, and care must therefore be taken in this time range. Moreover, there might be hidden fast $\gamma$-ray flares (i.e., short-term events on a scale of hours to minutes) that cannot be seen in our $\gamma$-ray light curve.

This suggests a connection between the $\gamma$-ray flares and the source opacity at (sub)mm wavelengths in the jet \citep[see e.g.,][for a similar case in the blazar \object{3C\,454.3}]{jorstad2013}.
\rva{Previous studies predicted a rapid decrease in the core shift with increasing radio frequencies \citep[see e.g.,][for a typical tendency of the core shift profiles]{dodson2017, lisakov2017, okino2022}. Our finding favors at least a small core shift at above 95\,GHz when $\gamma$-ray flares occur in the jet of PKS\,1424-418.
}

%
%A cross-correlation analysis is highly dependent on those fluxes in a flaring state \citep[e.g.,][]{kim2022}. As evident from Figure~\ref{fig:mmspidx}, the source becomes temporarily optically thick during the flaring periods at both $\gamma$-rays and (sub)mm-wavelengths (see Figure~\ref{fig:coreps}, for the relationship between the light curves). Hence, the weak delay of the significant correlation between the band 3 and $\gamma$-rays is likely to be attributed to those flatter $\alpha$ that occur occasionally. Interestingly, we find that the linear relationship in flux density between the band 3 and 7 data shown in Appendix~\ref{sec:allalma}, begin to deviate from the initial tight linear trend at an intermediate flux density level (i.e., F$_{\rm B7}$\,$\sim$\,3\,Jy). This further supports the above-mentioned interpretation on the variable source opacity at 95--345\,GHz.

%============================================================================================
\section{Discussion}
\label{sec:disccussion}

\subsection{Unusual long-term \rvc{mm}--\texorpdfstring{$\gamma$}{g}-ray correlation}
\label{sec:gunique}
In general, the detection of a significant \rvc{mm}--$\gamma$-ray correlation in blazars \er{has} been somewhat intermittent (for light curves spanning about 3--4\,yr overall), and \er{only} a small number of source samples \er{were} reported in previous statistical studies \cite[e.g.,][]{fuhrmann2014, maxm2014a, rama2015, rama2016}. It is even more \er{difficult} to search for these significant correlations in some of the blazars with extreme radio variability \citep[e.g., 0716$+$714,][]{kim2022}. 
This is partly because the previous works were done in the low-frequency radio bands (e.g., below 40\,GHz) where the radio emission arises from a region that lies farther away from the high-energy region \citep[see][for their Figure 3]{maxm2014a}.
We found a long-term tightly correlated (at $>$\,99.9\%) \rvc{mm}--$\gamma$-ray variability in \object{PKS\,1424$-$418} that is quite atypical. After the period of our datasets, the source became quiescent at both (sub)mm wavelengths and $\gamma$-rays for about one year (e.g., March\ 2020 to March\ 2021). Thus, we \er{assume} that this unique \rvc{mm}--$\gamma$-ray correlation spanned almost $\sim$10\,yr in the source.

Leptonic models (i.e., synchrotron and IC scattering) have been successful \er{in} describing most of \er{the} blazar SEDs \cite[e.g.,][]{bottcher2013, paliya2018}. This tight correlation between $\gamma$-ray and lower-energy bands strongly supports the IC processes for the $\gamma$-ray emission \citep[e.g.,][]{liodakis2019} via synchrotron-self Compton (SSC) and/or external Compton (EC) with seed photon fields from either the broad-line region (BLR) or a dusty torus (DT).
Previous SED studies indeed favored a combination of the SSC and EC processes for the observed $\gamma$-rays from \object{PKS\,1424$-$418} \citep{tavecc2013, buson2014, paliya2018, abhir2021}.
% but with different conclusions on the main source of the external photon field (i.e., BLR vs. DT) and the $\gamma$-ray dissipation zone (e.g., subpc vs. pc scales from the SMBH; but \er{in any case outside of the BLR region}).
%
However, further details about the mechanism are somewhat elusive.
%
% In our results, the $\gamma$-ray and band 7 emission regions seem co-located (i.e., \rv{near-zero} lag); including the band 3 emission during non-flaring periods.
%
% Such relationship is a typical diagnostic of the SSC $\gamma$-rays \citep[e.g.,][]{agudo2011b}.
Our results refer to a typical diagnostic of the SSC $\gamma$-rays \citep[e.g.,][]{agudo2011b}.
%: i.e., synchrotron$+$IC in the same emitting region.
The strong long-term \rvc{mm}--$\gamma$-ray correlation in \object{PKS\,1424$-$418} might be attributed to SSC of low-energy electrons, which leads to longer cooling timescales for the $\gamma$-rays \citep[e.g.,][]{bottcher2019}. However, the significant Compton dominance in this source \citep[e.g.,][]{paliya2018} indicates that strong external photon fields play an important role via EC (see also \citealt{marscher2010} and \citealt{macdonald2015} for a discussion of a slow jet sheath as a potential source of soft seed photons at \rvf{$\leq$}\,pc scales).
%
% \rva{The dominant source of the external seed photons may be DT at pc scales (e.g., $>$\,1\,pc from the central engine). It is generally assumed that the external photon energy density of BLR decreases rapidly between $\sim$0.1\,pc and 1\,pc scales, \er{while} the DT remains strong up to $\sim$10\,pc \citep[e.g.,][]{paliya2018}. In the meantime, a recent work \citet{agarwal2024} found the $\gamma$-ray absorption feature at above 10\,GeV in this source. This could be a strong evidence that the $\gamma$-rays originated from a region of influence of the BLR photons; see also \citet{marscher2010} and \citet{macdonald2015}, for discussion of a slow jet sheath as a potential source of soft seed photons at (sub)pc scales.}
% \er{For} the EC $\gamma$-rays, the dominant source of the external seed photons \er{may} be DT at pc scales (e.g., $>$\,1\,pc from the central engine) where the region is thought to be transparent at radio frequencies (see Section~\ref{sec:gorigin}). It is generally assumed that the external photon energy density of BLR decreases rapidly between $\sim$0.1\,pc and 1\,pc scales, \er{while} the DT remains strong up to $\sim$10\,pc (e.g., \citealt{paliya2018}; see also \citealt{marscher2010, macdonald2015}, for discussion of a slow jet sheath as a potential source of soft seed photons at pc scales).
% ; in general, 1\,pc corresponds to $10^{4}$\,R$_{\rm s}$, where R$_{\rm s}$ being Schwarzschild radius.

%\subsection{On the \texorpdfstring{$\gamma$}{g}-ray production site}
\subsection{\rva{Location of the \texorpdfstring{$\gamma$}{g}-ray production site}}
\label{sec:gorigin}
%
% Using $\gamma$-ray spectra of $\sim$100 bright FSRQs, \citet{costamante2018} searched for an absorption feature caused by the $\gamma$--$\gamma$ collision that is assumed to occur within the radius of BLR (i.e., typically $R_{\rm BLR}$ $\sim$ 0.1\,pc) at above a few tens of GeV energy bands. The authors found no strong evidence of the feature and concluded that the $\gamma$-rays \er{almost always originate outside} the BLR.
% This meets the jet model proposed by \citet{marscher2008}.
%
Based on the stable tight \rvc{mm}--$\gamma$-ray correlation over more than 8.5\,yr, it is straightforward to consider that the $\gamma$-ray origin is cospatial with the (sub)mm wave core region in the jet of \object{PKS\,1424$-$418} \citep[see also][for large delays between $\gamma$-ray and cm-wavelengths in the source]{vanzyl2015}.

The inner jet regions (e.g., $<$\,0.1--1\,pc) are thought to be opaque at radio frequencies \citep{marscher2008}, and radio emission therefore emerges beyond this region where the radio core appears. In blazars, the centimeter-wave core is assumed to be located at 7--12\,pc, for instance, from the central engine \citep{pushkarev2010, kramarenko2022}. Thus, the mm wave core is likely to be located at a distance scale smaller than this range, for example, $<$\,7\,pc, which is closer to the central engine at higher frequencies. 
This can be confirmed with mm VLBI observations. For instance, \citet{kim2023} presented a 3mm VLBI image of the jet of the blazar BL\,Lacertae ($z$ = 0.0686) observed by the Global mm VLBI Array (GMVA) on an angular resolution scale of tens of $\mu$as ($\sim$70\,$\mu$as). At \er{this redshift,}
%$z$ = 0.0686,
the image scale of the source is $\sim$1.3\,pc\,mas$^{-1}$.
%(assuming $H_{0}$ = 71 km~Mpc~s$^{-1}$, $\Omega_{\Lambda}$ = 0.73, and $\Omega_{m}$ = 0.27)
Using $M_{\rm BH}$\,$\sim$\,10$^{8.2}\,M_{\odot}$ \citep[e.g.,][]{cohen2014}, the angular resolution of 70\,$\mu$as can be converted into a projected distance of 0.091\,pc in the observer frame, which corresponds to $\sim$6000\,R$_{\rm s}$, where R$_{\rm s}$ is the Schwarzschild radius. With the jet viewing angles of 1--10$^{\circ}$ and assuming a conservative wide range for this blazar \citep[see e.g.,][]{marscher2008, weaver2022}, we can obtain a range of the deprojected distance that is $\sim$0.5--4.9\,pc in the source frame. These estimates might be considered as upper limits \er{for} the location of the (sub)mm
% \LEt{***because it occurs so often: abbreviated units don't take a plural s ("yr", not "yrs", and they are not hyphenated***}
wave core region in BL\,Lacertae.
It is worth noting that \citet{agarwal2024} recently found an absorption feature in $\gamma$-ray spectra of PKS\,1424-418 at above 10\,GeV \citep[but see also e.g.,][for the absence of such feature in the majority of FSRQs which is a more general case]{costamante2018}. This might be strong evidence that the $\gamma$-rays originated from a region of influence of the BLR photons (e.g., a typical radius of BLR $\sim$ 0.1\,pc). If this is the case, the presence of the $\gamma$-ray absorption might either imply the proximity of the (sub)mm wave core to the outer boundary of the BLR or indicated that an extended \rvc{structure} of the BLR lies farther downstream of the jet.
% Considering a typical radius of BLR (e.g., $R_{\rm BLR}$ $\sim$ 0.1\,pc), 

The SSA opacity ($\tau_{\rm opaq}$) of a jet region can be enhanced with a higher electron density, which meets the argument of moving shocks in the compact core regions \citep[e.g.,][]{kim2022}. \er{On the other hand}, \citet{sharma2022} presented the \er{effect} of \er{the} jet viewing angle ($\theta_{\rm jet}$) on $\tau_{\rm opaq}$ in a conical self-absorbed jet. The authors suggested that a spatial displacement between the radio cores with spectral hardening becomes more pronounced at smaller jet viewing angles. This is also consistent with an expression of the SSA opacity \citep[i.e.,][]{finke2019}, that is, $\tau_{\rm opaq}(\epsilon) \propto \delta^{(p+2)/2}$, where $\epsilon$ is approximately $\sim\,\nu / \rm (1.23\,\times\,10^{20})$\,Hz, $\delta$ is the Doppler factor, and $p$ is the power-law index of the electrons, $\alpha$ = ($p-1$)/2.
%
% Thus, there might be a change in $\theta_{\rm jet}$ around the (sub)mm-wave core region during the interaction between the core and \er{the} moving shocks (e.g., \er{towards} smaller $\theta_{\rm jet}$).
Thus, there might be
a change in $\theta_{\rm jet}$ near the radio core region
and/or
a core shift in the (sub)mm waveband
during the interaction between the core and \er{the} moving shocks (e.g., \er{toward} smaller $\theta_{\rm jet}$).

% Our observational results favor the above-mentioned scenario.
% Given the stable, tight \rvc{mm}--$\gamma$-ray correlation over more than 8.5\,yr, it is straightforward to consider that the $\gamma$-ray origin is co-spatial with the (sub)mm-wave core region in the jet of \object{PKS\,1424$-$418} \citep[see also][for large delays between $\gamma$-ray and cm-wavelengths in the source]{vanzyl2015}.
% An exception can \er{only occur if} the source is flaring as we find a small time delay of $\sim$3\,days at band 3; this indicates a small displacement between the $\gamma$-ray site and the 95\,GHz core. However, we expect that even such displacement disappears completely at higher radio frequencies (e.g., $\geq$\,345\,GHz).

%\subsection{\rva{Presence of a time lag in the LCCF curves between the radio and \texorpdfstring{$\gamma$}{g}-ray light curves?}}
%\subsection{\rva{Potential effects of a core shift at above 95\,GHz to the \texorpdfstring{$\gamma$}{g}-ray origin}}
\subsection{\rva{Potential effects of a core shift to the \texorpdfstring{$\gamma$}{g}-ray origin}}
\label{sec:smaldel}
\rva{
The coincidence of the spectral hardening with the $\gamma$-ray flares suggests that a gap lies between $\tau_{\rm \gamma-B3}$ and $\tau_{\rm \gamma-B7}$ that was caused by a core shift between 95\,GHz and 345\,GHz in the jet. To explore this possibility, we examined the LCCF curves of the $\gamma$-rays with the band 3 and band 7 datasets (Appendix~\ref{sec:gausfit}).
The LCCF curve of the band 7 peaks at two positions: first, at $\tau$\,=\,$-$16\,days (LCCF\,$\sim$\,0.78$\pm$0.15), and then, at $\tau$\,=\,0 (0.77$\pm$0.12), where negative $\tau_{\rm \gamma-radio}$ means ``$\gamma$-ray leading''.
To better estimate probable locations of the LCCF peaks for both $\tau_{\rm \gamma-B3}$ and $\tau_{\rm \gamma-B7}$, we fit a Gaussian model to the data. The fits estimate $\tau_{\rm \gamma-B3}$\,=\,$-$2.8$\pm$2.0\,days and $\tau_{\rm \gamma-B7}$\,=\,$-$0.8$\pm$2.6\,days. 
This result is consistent with the model of the core shift (i.e., the higher the frequencies, the more upstream the regions) and suggests that the $\gamma$-rays are almost simultaneous with the band 7 flux densities, but precede the band 3 flux densities by a few days. A cross-correlation analysis is highly dependent on these flux densities in a flaring state \citep[e.g.,][]{kim2022}. As evident from Figure~\ref{fig:mmspidx}, we consider that the source becomes temporarily optically thick at (sub)mm wavelengths during the $\gamma$-ray flaring periods\rvc{. We also note that the photon flux density of blazar $\gamma$-ray flares generally tends to be higher with shorter binning intervals}.
Hence, we cannot rule out the possibility that regions with flatter $\alpha_{\rm radio}$ could cause the weak delay between the band 3 and $\gamma$-ray emission.
\rvc{This means that the band 3 emission comes days after the $\gamma$-rays and band 7 emission.
}
We also note that the linear relation between the band 3 and 7 emission shown in Appendix~\ref{sec:allalma} begins to deviate from an initially tight linear trend at around F$_{\rm B7}$\,$\sim$\,2.8\,Jy. This further supports the above-mentioned interpretation of the variable source opacity at 95--345\,GHz.
In this regard, we suggest that the $\gamma$-ray site and the 95\,GHz core position are only slightly displaced when the source is flaring at $\gamma$-rays. However, we expect that even this displacement disappears completely at higher radio frequencies (e.g., $\geq$\,345\,GHz) \rvc{because the radio core becomes transparent with increasing radio frequency}.
}

% interpretation of the delays
The distance ($\Delta d_{\rm \gamma-radio}$) between the mm and $\gamma$-ray emitting regions can be estimated by $\Delta d_{\rm \gamma-radio}$ = $\beta_{\rm app}\,c\,\tau_{\rm \gamma-radio}\,\textup{sin}^{-1}\theta_{\rm jet}\,(1+z)^{-1}$, where $\beta_{\rm app}$ is the apparent jet speed, $c$ is the speed of light, $\tau_{\rm \gamma-radio}$ is the delay between radio and $\gamma$-ray light curves, $\theta_{\rm jet}$ is the jet viewing angle, and $z$ is the source redshift. Assuming $\beta_{\rm app}$ = 9 (i.e., a median value for FSRQs; \citealt{weaver2022}) and $\theta_{\rm jet}$ = 1--3$^{\circ}$ \citep[e.g.,][]{paliya2018, weaver2022}, we find a deprojected distance of $\Delta d_{\rm \gamma-B3}$\,$\sim$\,0.16\,pc with $\theta_{\rm jet}$ = 3$^{\circ}$ for $\tau_{\rm \gamma-B3}$ = 2.8\,days (Figure~\ref{fig:lccfmain}) and $\Delta d_{\rm \gamma-B3}$\,$\sim$\,0.48\,pc with $\theta_{\rm jet}$ = 1$^{\circ}$. This indicates that the $\gamma$-ray production site is located in a region upstream of the 95\,GHz core by about 0.16--0.48\,pc; but if $\beta_{\rm app}$ is much higher \citep[e.g., $\sim$40;][]{weaver2022}, then the distance range for the same $\theta_{\rm jet}$ = 1--3$^{\circ}$ can be $\Delta d_{\rm \gamma-B3}$\,$\sim$\,0.71--2.14\,pc. For the $\tau_{\rm \gamma-B7}$ estimate of 0.8\,day, it is basically a zero-lag within the uncertainty. If there is at least a small shift in the position of the 345\,GHz core and we consider the $\tau_{\rm \gamma-B7}$ of 0.8\,day as the corresponding shift, however, the distance between the 345\,GHz core and the $\gamma$-ray site could be $\Delta d_{\rm \gamma-B7}$\,$\sim$\,0.05--0.14\,pc for the same $\theta_{\rm jet}$ range above at $\beta_{\rm app}$\,=\,9\rvc{. This distance scale is comparable to the radius of the BLR estimated by \citet{tavecc2013}}.

% in VLBI images
A \er{simple} way to confirm the above predictions is \er{to perform} multiband (quasi-)simultaneous VLBI observations of the source at (sub)mm wavelengths to measure the core shift. At $z$ = 1.52, the luminosity distance of \object{PKS\,1424$-$418} is $\sim$11192\,Mpc, and this corresponds to an image scale of 8.54\,pc/mas.
%(\er{using} the same cosmological parameters above)
We find projected distances of $\sim$0.02\,pc or $\sim$2.5\,$\mu$as for the $\Delta d_{\rm \gamma-B3}$ values, and if present, $\sim$0.006\,pc or $\sim$0.7\,$\mu$as for the $\Delta d_{\rm \gamma-B7}$ values. This suggests that the level of the core shift between the 95\,GHz and 345\,GHz cores is $\sim$1.8\,$\mu$as, which \er{may indeed} be present in these high-frequency core regions.

%\subsection{\rva{Location of the \texorpdfstring{$\gamma$}{g}-ray production site}}
%\subsection{\rva{On the (sub)millimeter spectral hardening}}\label{sec:alphavar}
\subsection{\rva{Physical scenario for the production of the \texorpdfstring{$\gamma$}{g}-ray flares}}\label{sec:alphavar}

% variations in the core opacity
From decades of VLBI studies of blazar jets, it is \er{clear} that the radio core is a dominant source of the observed radio emission from the jets, especially when a strong disturbance (or a moving shock) \er{passes} through the core region \citep[e.g.,][]{marscher2008}. This is particularly true at higher radio frequencies, \er{since} there should be no significant contributions \er{from} the extended jet regions to the observed radio emission \citep{marscher1985}. Higher-energy electrons last much shorter than lower-energy ones regarding their energetic lifetimes. We find \er{remarkable} variations in the jet opacity at 95--345\,GHz \er{that coincide} with the $\gamma$-ray flaring periods. \citet{plavin2019} found a core-shift variability at low frequencies (i.e., 2--8\,GHz) with large source samples. This \er{may} be \er{related to} our findings. As \er{suggested by} the authors, the passage of a moving shock through the core region could lead to a displacement in the core position by pushing it downstream. If this holds in our case, the following scenario can be suggested: (1) The radio core was initially optically thin at above 95\,GHz \rvf{(i.e., $\leq$\,3.5\,mm).
% Alternatively, for instance, $\leq$\,3.5\,mm,
(2) A} moving shock causes core shifts at (sub)mm wavelengths by temporarily increasing the jet opacity \rvc{\citep[e.g.,][]{jorstad2013, wehrle2016, lisakov2017}}. (3) As the shock leaves the region, the core restores its initial condition (i.e., opacity and position).
\section{Summary}
\label{sec:fin}
We have \er{studied the time} correlation between an \rvc{mm wave} and $\gamma$-ray emission in the blazar \object{PKS\,1424$-$418} by using ALMA (sub)\rvc{mm} radio light curves at 90--350\,GHz and a LAT $\gamma$-ray light curve at 0.1--200\,GeV.
%
% \er{Due} to the cadence of data sampling, we mainly used the ALMA band 3 ($\sim$95\,GHz) \& 7 ($\sim$345\,GHz) data \er{for} the correlation analyses.
Because of the cadence of the data sampling, we mainly used ALMA band 3 ($\sim$95\,GHz) data with the band 6 and band 7 (235\,GHz and 345\,GHz, respectively) data as supporting material \er{for} the correlation analyses.
Interestingly, the \rvc{band 3} and $\gamma$-ray light curves \er{already appear to be very similar to each other visually,} 
%look very similar to each other already from visual inspection 
and we \er{robustly} confirmed this with detailed statistical analyses.
We find a highly significant positive \rvc{mm}--$\gamma$-ray correlation over the
% whole
\rv{complete time period of}
8.5\,yr, which means that this blazar is the best case thus far of a long-term close connection between mm wave and $\gamma$-ray emission (A. Marscher, private communication).
This long-term \er{strong} correlation \er{is} typically observed between optical and $\gamma$-ray emission in blazars. 
%
% This suggests that the radio emission at higher frequencies (e.g., $>$\,90\,GHz) almost always originates from a region of the high-energy emission in this source. 
\rva{This suggests that the $\gamma$-ray emission almost always originates in 
the (sub)mm wave radio core region (i.e., $>$\,90\,GHz) in this source.}
\rva{If the core shift indeed occurs at the high radio frequencies with \rvc{strong $\gamma$-ray flares}, however, an exception can occur when a moving shock passes through the core region. This leads to an increase in the SSA opacity and thus to a displacement between the $\gamma$-ray site and the \rvc{mm wave} core.}
%
% The evolution of the spectral indices between the ALMA band 3 \& 7 flux densities \er{supports} this scenario.
%
 \er{A} flare-induced core shift at (sub)\rvc{mm} wavelengths like this could be confirmed by using a multiband VLBI campaign \er{involving} extremely high-resolution mm VLBI arrays
% \citep[see e.g.,][for a review on mm-VLBI]{boccardi2017}
\rv{\citep[see also review by][]{boccardi2017}}
such as GMVA and the Event Horizon Telescope (EHT).
It is worthwhile to note that our results are based on a long-term global trend, and there might be other $\gamma$-ray events in the source that deviate from our conclusion. 
%This can be examined by a monitoring campaign at submillimeter wavelengths (i.e., $>$\,300\,GHz) with better sampling cadence (e.g., $<$\,5\,days).
A monitoring campaign at shorter \rvc{mm} wavelengths (e.g., $<$\,3.5\,mm) with \er{a} better sampling cadence (i.e., $\leq$\,3\,days) would facilitate the \er{investigation} of these cases (e.g., orphan $\gamma$-ray events).

\begin{acknowledgements}
% general
% (AFTER 1st referee report) We thank the anonymous referee for comprehensive and constructive feedback that improved this paper.

% Alan, Svetlana, Yuri
We thank \rve{Alan Marscher \&} Svetlana Jorstad (Boston University), Yuri Kovalev (MPIfR), \rve{and} Abhishek Desai \rve{\&} Bindu Rani (GSFC) for useful comments on the manuscript.
% fermi-lat
%This work makes use of public Fermi data obtained from \textit{Fermi} Science Support Center (FSSC). 
The \textsl{Fermi} LAT Collaboration acknowledges generous ongoing support
from a number of agencies and institutes that have supported both the
development and the operation of the LAT as well as scientific data analysis.
These include the National Aeronautics and Space Administration and the
Department of Energy in the United States, the Commissariat \`a l'Energie Atomique
and the Centre National de la Recherche Scientifique / Institut National de Physique
Nucl\'eaire et de Physique des Particules in France, the Agenzia Spaziale Italiana
and the Istituto Nazionale di Fisica Nucleare in Italy, the Ministry of Education,
Culture, Sports, Science and Technology (MEXT), High Energy Accelerator Research
Organization (KEK) and Japan Aerospace Exploration Agency (JAXA) in Japan, and
the K.~A.~Wallenberg Foundation, the Swedish Research Council and the
Swedish National Space Board in Sweden.
 
Additional support for science analysis during the operations phase is gratefully
acknowledged from the Istituto Nazionale di Astrofisica in Italy and the Centre
National d'\'Etudes Spatiales in France. This work performed in part under DOE
Contract DE-AC02-76SF00515.
% alma
This paper makes use of the following ALMA data: ADS/JAO.ALMA\#2011.0.00001.CAL. ALMA is a partnership of ESO (representing its member states), NSF (USA) and NINS (Japan), together with NRC (Canada), MOST and ASIAA (Taiwan), and KASI (Republic of Korea), in cooperation with the Republic of Chile. The Joint ALMA Observatory is operated by ESO, AUI/NRAO and NAOJ.
% -- Grants --
% mine
This research was supported by Basic Science Research Program through the National Research Foundation of Korea (NRF) funded by the Ministry of Education (\rve{NRF-}2022R1A6A3A03069095).
% m2f
\er{This publication is part of the M2FINDERS project which
has received funding from the European Research Council (ERC) under the
European Union’s Horizon 2020 Research and Innovation Programme (grant
agreement No 101018682).}
% MK/FR
M.K. and F.R. acknowledge funding by the Deutsche Forschungsgemeinschaft (DFG, German Research Foundation) - grant 434448349.
% etc.
%\newline\textbf{Any missing grants!}

\end{acknowledgements}


\begin{thebibliography}{}

% \bibitem[Abdo et al.(2011)]{abdo2011} Abdo, A. A., Ackermann, M., Ajello, M., et al. 2011, ApJ, 726, 43

\bibitem[Abdollahi et al.(2020)]{abdollahi2020} Abdollahi, S., Acero, F., Ackermann, M., et al. 2020, ApJS, 247, 33

\bibitem[Abdollahi et al.(2023)]{abdollahi2023} \rvc{Abdollahi, S., Ajello, M., Baldini, L., et al. 2023, ApJS, 265, 31}

\bibitem[Abhir et al.(2021)]{abhir2021} Abhir, J., Joseph, J., Patel, S. R., et al. 2021, MNRAS, 501, 2504

\bibitem[Agarwal et al.(2024)]{agarwal2024} Agarwal, S., Shukla, A., Mannheim, K., et al. 2024, ApJL, 968, L1% in press (arXiv:2405.09612)

\bibitem[Agudo et al.(2011a)]{agudo2011a} Agudo, I., Jorstad, S. G., Marscher, A. P., et al. 2011a, ApJL, 726, L13

\bibitem[Agudo et al.(2011b)]{agudo2011b} Agudo, I., Marscher, A. P., Jorstad, S. G., et al. 2011b, ApJL, 735, L10

\bibitem[Ajello et al.(2020)]{ajello2020} Ajello, M., Angioni, R., Axelsson, M., et al. 2020, ApJ, 892, 105

\bibitem[Atwood et al.(2009)]{atwood2009} Atwood, W. B., Abdo, A. A., Ackermann, M., et al. 2009, ApJ, 697, 1071

\bibitem[Ballet et al.(2023)]{ballet2023} Ballet, J., Bruel, P., Burnett, T. H., et al. 2023, arXiv:2307.12546

\bibitem[Benke et al.(2024)]{benke2024} Benke, P., R\"{o}sch, F., Ros, E., et al. 2024, A\&A, 681, A69

\bibitem[Blandford et al.(2019)]{blandford2019} Blandford, R., Meier, D., \& Readhead, A. 2019, ARA\&A, 57, 467

\bibitem[Boccardi et al.(2017)]{boccardi2017} Boccardi, B., Krichbaum, T. P., Ros, E., et al. 2017, A\&ARv, 25, 4

\bibitem[Bonato et al.(2018)]{bonato2018} Bonato, M., Liuzzo, E., Giannetti, A., et al. 2018, MNRAS, 478, 1512

\bibitem[B\"{o}ttcher(2007)]{bottcher2007} B\"{o}ttcher, M. 2007, Ap\&\&SS, 309, 95

\bibitem[B\"{o}ttcher et al.(2013)]{bottcher2013} B\"{o}ttcher, M., Reimer, A., Sweeney, K., et al. 2013, ApJ, 768, 54

\bibitem[B\"{o}ttcher et al.(2019)]{bottcher2019} B\"{o}ttcher, M., \& Baring, M. G. 2019, ApJ, 887, 133

\bibitem[Buson et al.(2014)]{buson2014} Buson, S., Longo, F., Larsson, S., et al. 2014, A\&A, 569, A40

\bibitem[Chamani et al.(2023)]{chamani2023} Chamani, W., Savolainen, T., Ros, E., et al. 2023, A\&A, 672, A130

\bibitem[Cheung et al.(2007)]{cheung2007} \rvc{Cheung, C. C., Harris, D. E., Stawarz, \L{}. 2007, ApJL, 663, L65}

\bibitem[Cohen et al.(2014)]{cohen2014} Cohen, M. H., Meier, D. L., Arshakian, T. G., et al. 2014, ApJ, 787, 151

\bibitem[Connolly(2015)]{conoly2015} Connolly, S. D. 2015, arXiv:1503.06676

\bibitem[Costamante et al.(2018)]{costamante2018} Costamante, L., Cutini, S., Tosti, G., et al. 2018, MNRAS, 477, 4749

\bibitem[Dar \& Laor(1997)]{dar1997} Dar, A., \& Laor, A. 1997, ApJ, 478, L5

\bibitem[Dodson et al.(2017)]{dodson2017} Dodson, R., Rioja, M. J., Molina, S. N., et al. 2017, ApJ, 834, 177

\bibitem[Edelson \& Krolik(1988)]{edelson1988} Edelson, R. A., \& Krolik, J. H. 1988, ApJ, 333, 646

\bibitem[Emmanoulopoulos et al.(2013)]{emma2013} Emmanoulopoulos, D., McHardy, I. M., \& Papadakis, I. E. 2013, MNRAS, 433, 907

\bibitem[Finke(2019)]{finke2019} Finke, J. D. 2019, ApJ, 870, 28

\bibitem[Fuhrmann et al.(2014)]{fuhrmann2014} Fuhrmann, L., Larsson, S., Chiang, J. 2014, MNRAS, 441, 1899

\bibitem[Hodgson et al.(2018)]{hodgson2018} \rvc{Hodgson, J. A., Rani, B., Lee, S. -S., et al. 2018, MNRAS, 475, 368}

\bibitem[Jorstad et al.(2001a)]{jorstad2001a} Jorstad, S. G., Marscher, A. P., Mattox, J. R., et al. 2001a, ApJS, 134, 181

\bibitem[Jorstad et al.(2001b)]{jorstad2001b} Jorstad, S. G., Marscher, A. P., Mattox, J. R., et al. 2001b, ApJ, 556, 738

\bibitem[Jorstad et al.(2013)]{jorstad2013} Jorstad, S. G., Marscher, A. P., Smith, P. S., et al. 2013, ApJ, 773, 147

\bibitem[Kadler et al.(2016)]{kadler2016} Kadler, M., Krau{\ss}, F., Mannheim, K. 2016, NatPh, 12, 807

\bibitem[Keenan et al.(2021)]{keenan2021} Keenan, M., Meyer, E. T., Georganopoulos, M., et al. 2021, MNRAS, 505, 4726

\bibitem[Kim et al.(2018)]{kim2018} Kim, D. -W., Trippe, S., Lee, S. -S. 2018, MNRAS, 480, 2324

\bibitem[Kim et al.(2020)]{kim2020} Kim, D. -W., Trippe, S., \& Kravchenko, E. V. 2020, A\&A, 636, A62

\bibitem[Kim et al.(2022)]{kim2022} Kim, D. -W., Kravchenko, E. V., Kutkin, A. M., et al. 2022, ApJ, 925, 64

\bibitem[Kim et al.(2023)]{kim2023} Kim, D. -W., Janssen, M., Krichbaum, T. P., et al. 2023, A\&A, 680, L3

\bibitem[Kovalev et al.(2009)]{kovalev2009} Kovalev, Y. Y., Aller, H. D., Aller, M. F., et al. 2009, ApJL, 696, L17

\bibitem[Kramarenko et al.(2022)]{kramarenko2022} Kramarenko, I. G., Pushkarev, A. B., Kovalev, Y. Y., et al. 2022, MNRAS, 510, 469

%  \bibitem[Kudryavtseva et al.(2011)]{kudryavtseva2011} Kudryavtseva, N. A., Gabuzda, D. C., Aller, M. F., et al. 2011, MNRAS, 415, 1631

\bibitem[Lee et al.(2008)]{lee2008} Lee, S. -S., Lobanov, A. P., Krichbaum. T. P., et al. 2008, AJ, 136, 159

\bibitem[Le\'{o}n-Tavares et al.(2011)]{leon2011} Le\'{o}n-Tavares, J., Valtaoja, E., Tornikoski, M., et al. 2011, A\&A, 532, A146

\bibitem[Liodakis et al.(2019)]{liodakis2019} Liodakis, I., Romani, R. W., Filippenko, A. V., et al. 2019, ApJ, 880, 32

\bibitem[Liodakis et al.(2020)]{liodakis2020} \rvc{Liodakis, I., Blinov, D., Jorstad, S. G., et al. 2020, ApJ, 902, 61}

\bibitem[Lisakov et al.(2017)]{lisakov2017} Lisakov, M. M., Kovalev, Y. Y., Savolainen, T., et al. 2017, MNRAS, 468, 4478

\bibitem[Lister et al.(2009)]{lister2009} Lister, M. L., Aller, H. D., Aller, M. F., et al. 2009, AJ, 137, 3718

\bibitem[Lobanov(1998)]{lobanov1998} Lobanov, A. P. 1998, A\&A, 330, 79

\bibitem[MacDonald et al.(2015)]{macdonald2015} MacDonald, N. R., Marscher, A. P., Jorstad, S. G., et al. 2015, ApJ, 804, 111

\bibitem[MAGIC Collaboration et al.(2018)]{magic2018} MAGIC Collaboration, Ahnen, M. L., Ansoldi, S., et al. 2018, A\&A, 619, A45

\bibitem[Marscher \& Gear(1985)]{marscher1985} Marscher, A. P., \& Gear, W. K. 1985, ApJ, 298, 114

\bibitem[Marscher et al.(2008)]{marscher2008} Marscher, A. P., Jorstad, S. G., D`Arcangelo, F. D., et al. 2008, Nature, 452, 966

\bibitem[Marscher et al.(2010)]{marscher2010} Marscher, A. P., Jorstad, S. G., Larionov, V. M., et al. 2010, ApJL, 710, L126

\bibitem[Marscher(2016)]{marscher2016} Marscher, A. P. 2016, Galax, 4, 37

\bibitem[Max-Moerbeck et al.(2014a)]{maxm2014a} Max-Moerbeck, W., Hovatta, T., Richards, J. L., et al. 2014a, MNRAS, 445, 428

\bibitem[Max-Moerbeck et al.(2014b)]{maxm2014b} Max-Moerbeck, W., Richards, J. L., Hovatta, T., et al. 2014b, MNRAS, 445, 437

%  \bibitem[Meyer et al.(2019)]{meyer2019} Meyer, M., Scargle, J. D., \& Blandford, R. D. 2019, ApJ, 877, 39

\bibitem[Nair et al.(2019)]{nair2019} Nair, D. G., Lobanov, A. P., Krichbaum, T. P., et al. 2019, A\&A, 622, A92

%  \bibitem[Ojha et al.(2010)]{ojha2010} Ojha, R., Kadler, M., B\"{o}ck, M., et al. 2010, A\&A, 519, A45

\bibitem[Okino et al.(2022)]{okino2022} Okino, H., Akiyama, K., Asada, K., et al. 2022, ApJ, 940, 65

\bibitem[Paliya et al.(2018)]{paliya2018} Paliya, V. S., Zhang, H., B\"{o}ttcher, M., et al. 2018, ApJ, 863, 98

% \bibitem[Paraschos et al.(2023)]{paraschos2023} Paraschos, G. F., Mpisketzis, V., Kim, J. -Y. 2023, A\&A, 669, A32

\bibitem[Plavin et al.(2019)]{plavin2019} Plavin, A. V., Kovalev, Y. Y., Pushkarev, A. B., et al. 2019, MNRAS, 485, 1822

\bibitem[Pushkarev et al.(2010)]{pushkarev2010} Pushkarev, A. B., Kovalev, Y. Y., \& Lister, M. L. 2010, ApJL, 722, L7

\bibitem[Ramakrishnan et al.(2015)]{rama2015} Ramakrishnan, V., Hovatta, T., Nieppola, E. 2015, MNRAS, 452, 1280

\bibitem[Ramakrishnan et al.(2016)]{rama2016} Ramakrishnan, V., Hovatta, T., Tornikoski, M. 2016, MNRAS, 456, 171

\bibitem[Rani et al.(2018)]{rani2018} Rani, B., Jorstad, S. G., Marscher, A. P., et al. 2018, ApJ, 858, 80

\bibitem[R\"{o}der et al.(2024)]{roder2024} R\"{o}der, J., Ros, E., Schinzel, F. K., et al. 2024, A\&A, 684, A211

\bibitem[Sharma et al.(2022)]{sharma2022} Sharma, R., Massi, M., \& Torricelli-Ciamponi, G. 2022, A\&A, 660, A58

\bibitem[Tavecchio et al.(2013)]{tavecc2013} Tavecchio, F., Pacciani, L., Donnarumma, I., et al. 2013, MNRAS, 435, L24

\bibitem[van Zyl \& Gaylard(2015)]{vanzyl2015} van Zyl, P. V., \& Gaylard, M. J. 2015, MmSAI, 86, 36

\bibitem[Weaver et al.(2022)]{weaver2022} Weaver, Z. R., Jorstad, S. G., Marscher, A. P., et al. 2022, ApJS, 260, 12

\bibitem[Wehrle et al.(2012)]{wehrle2012} Wehrle, A. E., Marscher, A. P., Jorstad, S. G., et al. 2012, ApJ, 758, 72

\bibitem[Wehrle et al.(2016)]{wehrle2016} Wehrle, A. E., Grupe, D., Jorstad, S. G., et al. 2016, ApJ, 816, 53

\bibitem[Welsh(1999)]{welsh1999} Welsh, W. F. 1999, PASP, 111, 1347

\bibitem[White et al.(1988)]{white1988} White, G. L., Jauncey, D. L., Savage, A., et al. 1988, ApJ, 327, 561


\end{thebibliography}

%\appendix
\begin{appendix}

\section{The ALMA band 3, 6, and 7 data}
\begin{figure}[!htbp]
\centering
\includegraphics[angle=0, width=\columnwidth, keepaspectratio]{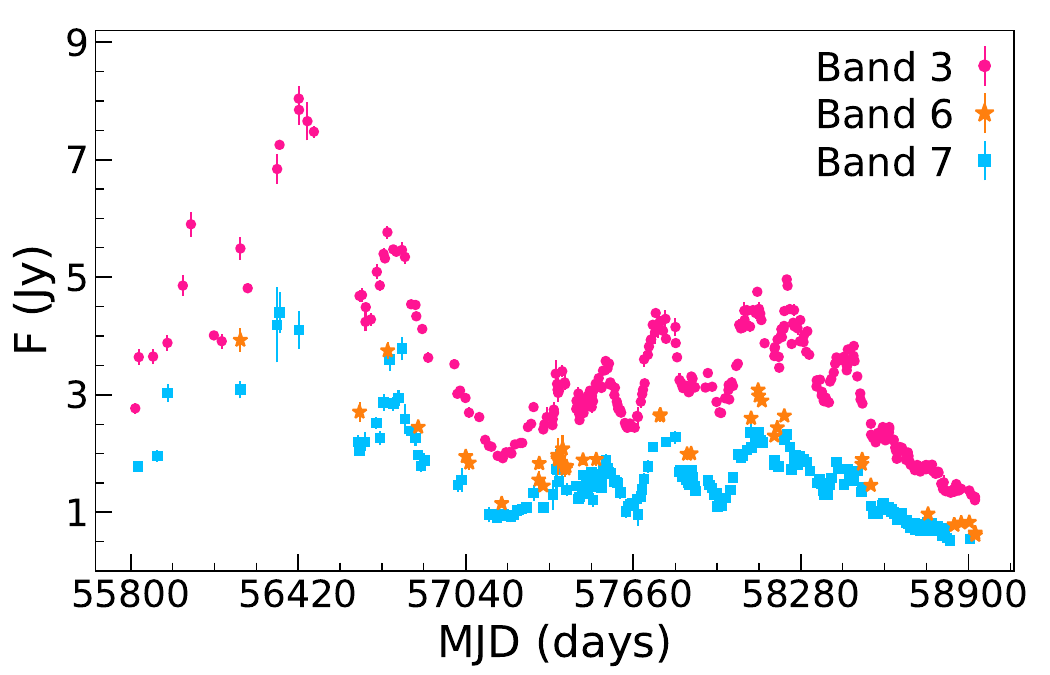} \
\includegraphics[angle=0, width=\columnwidth, keepaspectratio]{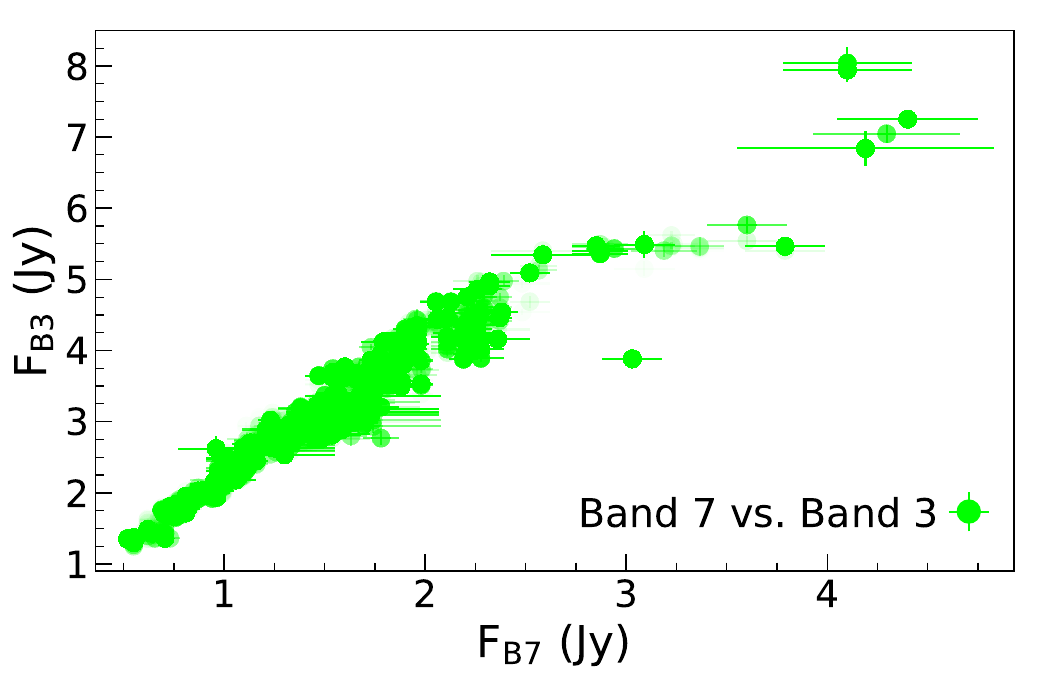} \
\includegraphics[angle=0, width=\columnwidth, keepaspectratio]{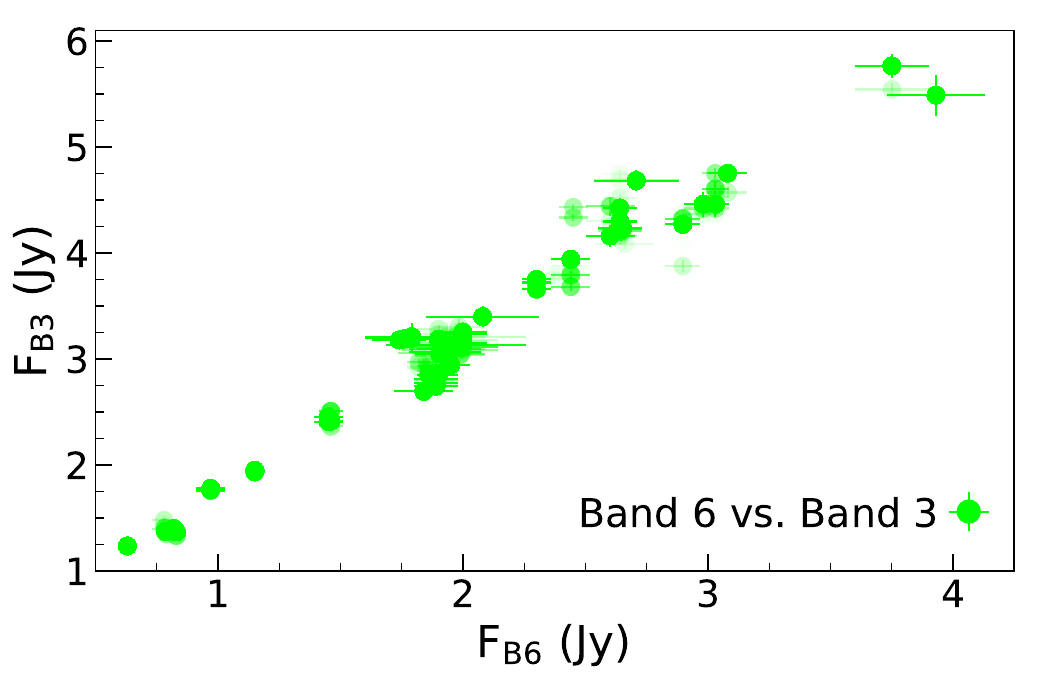} \
\includegraphics[angle=0, width=\columnwidth, keepaspectratio]{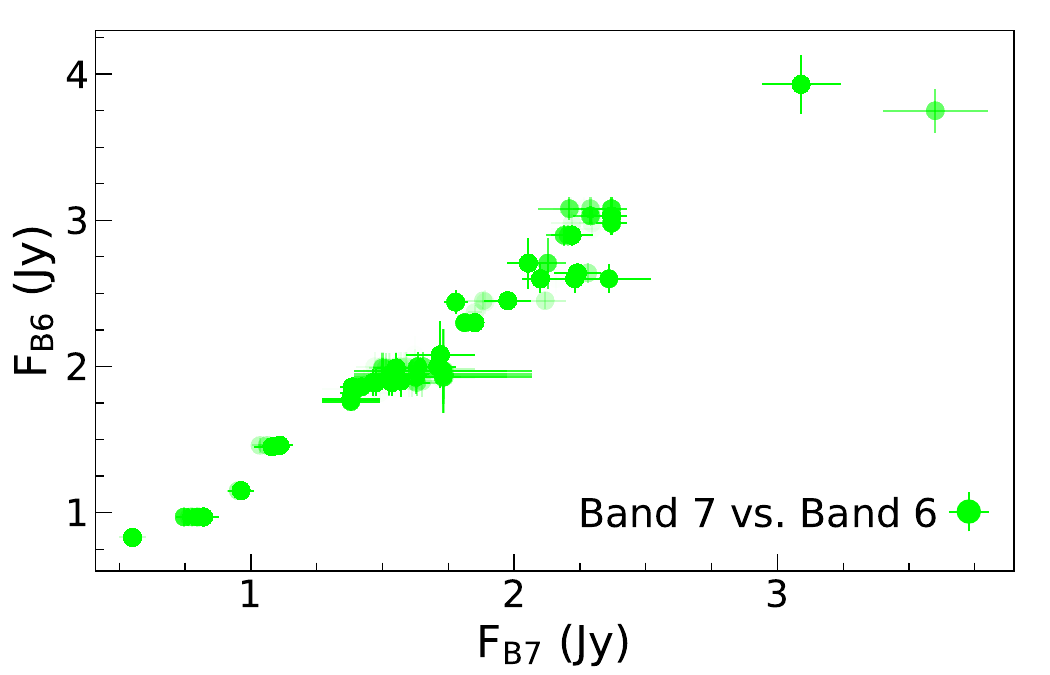}
\caption{
From top to bottom: ALMA light curves of \object{PKS\,1424$-$418} at band 3 (95\,GHz), band 6 (235\,GHz), and band 7 (345\,GHz) and plots of the flux--flux scatter ($\sim$930 samples for B7--B3 \& $\sim$830 samples for the others) for each pair of the ALMA bands.
}
\label{fig:almabands}
\end{figure}

%\section{Pearson correlations between band 3/6/7}
\label{sec:allalma}
%
% Although \er{the} data sampling is worse than \er{at} ALMA band 3 (B3),
\rv{Although the data sampling is best in the band 3 (B3),}
we were able to obtain useful radio light curves of the source at other ALMA bands: band 6 (B6; $\sim$235\,GHz) and band 7 (B7; $\sim$345\,GHz). Their sampling intervals are, on average, $\sim$66\,days with a median value of $\sim$25\,days at band 6 and $\sim$14\,days with a median value of $\sim$8\,days at band 7. Figure~\ref{fig:almabands} shows all the ALMA light curves of the source.
For every pair of the two time series (i.e., B3--B6, B6--B7, and B7--B3), we calculated \er{the} Pearson correlation coefficients following the manner described in Appendix~\ref{sec:panal}. The Bin$_{\rm eq}$ ranges were set to be 3--30\,days for the B7--B3 pair (i.e., $\sim$930 flux--flux sets) and 3--15\,days for the other two pairs with band 6 (i.e., $\sim$830 flux--flux sets). This is due to the poor sampling of the band 6 data which is significantly worse (e.g., 4--5 times) than the other two bands.
The Bin$_{\rm eq}$ ranges correspond to NBin$_{\rm eq}$ of $\sim$1040 to $\sim$110 for B7--B3 and $\sim$1040 to $\sim$210 for the others.
%37 -- $\sim$930 samples
%In [366]: nranges
%Out[366]: [105, 1038]
%36 -- $\sim$830 samples
%In [368]: nranges
%Out[368]: [209, 1039]
%67 -- $\sim$830 samples
%In [370]: nranges
%Out[370]: [208, 1035]
%
The results on the three pairs are shown in Figure~\ref{fig:almabands}. We find the averages of r$_{\rm P}$ as follow: 0.967$\pm$0.004 (with 1\,$\sigma$ uncertainty) for B3--B7, 0.987$\pm$0.002 for B3--B6, and 0.988$\pm$0.005 for B7--B6. 
% time stamp
%We investigated linear relationships between the radio light curves by using Pearson correlation. For every pair of the two time series (i.e., B3--B6, B6--B7, and B7--B3), we determined a time range where it begins and ends at minimum and maximum values in the time domain between two light curves, respectively.
% data selection
%Then, we split the time range into a number of equal-sized bins (N$_{\rm Tbins}$). Since the fluxes were sampled unevenly and randomly distributed, we tested this experiment with N$_{\rm Tbins}$ = 30 to 300 (thus total 271 sets of the correlation estimates for each of the light curve pairs). This corresponds to the equal-sized bins (Tbins) of $\sim$100\,days to $\sim$10\,days.
% the calculation
%In each of the Tbins, we calculated the average of all fluxes belonging to the Tbin and used it as a typical flux density of the Tbin. Once this process is done for two time series separately, we only selected Tbins overlapping with each other (e.g., B3 vs. B7) and calculated a Pearson correlation coefficient (r$_{\rm P}$) with those selected fluxes.

%# B3 (btw. 91.5 GHz & 103.5 GHz)
%fB3 = 95e9
%# B6
%fB6 = 233e9
%# B7
%fB7 = 345e9

\section{Testing Pearson correlation}
\label{sec:panal}
%
% time stamp
We investigated linear relationships between the light curves by using the Pearson correlation. For \er{each} pair of two time series (e.g., \rvc{mm} \& $\gamma$-ray), we determined a time range of the data where it begins at \er{the} minimum of the two time series and ends at \er{the} maximum of the two time series.
% where it begins and ends at minimum and maximum values in the time domain between the two light curves, respectively.
% data selection
\er{The full time range was divided into equal-sized bins (Bin$_{\rm eq}$) to account for the uneven and random distribution of the data. }
%Then, we split this full time range into a number of equal-sized bin (Bin$_{\rm eq}$). Since the data were sampled unevenly and randomly distributed. 
\er{To ensure accuracy, this test was repeated in a range of different \rvc{sizes} of Bin$_{\rm eq}$, taking into account the data sampling and source variability}; \rvc{NBin$_{\rm eq}$ is the total number of Bin$_{\rm eq}$ in a single test run.}
%we iterated this test on a broad range of the number of Bin$_{\rm eq}$ (NBin$_{\rm eq}$); the NBin$_{\rm eq}$ range was set considering data sampling and source variability
% \rmv{(see e.g., Appendix~\ref{sec:allalma})}.
% {\color{green} \textst{(see e.g., Appendix~{\color{blue} A})}}.
% \textst{(see e.g., Appendix~{\color{blue} A})}.
% the calculation
We calculated an average of all \rvc{the} flux densities belonging to each \rvc{of the} Bin$_{\rm eq}$ and used it as a typical flux density \rvc{for each of them}. Once this process is done throughout all \rvc{NBin$_{\rm eq}$} on the two light curves separately, we only selected Bin$_{\rm eq}$ overlapping with each other and calculated a Pearson correlation coefficient (r$_{\rm P}$) by using these selected local-mean flux densities.

\section{Cross-correlation analysis}
\label{sec:ccf1}
%
% choice of the method
The Discrete Correlation Function \citep[DCF;][]{edelson1988} has been widely employed in the analysis of time-correlation between two light curves. To avoid the normalization issues of DCF \citep[i.e.,][]{maxm2014b}, however, we employed the Local Cross-Correlation Function \citep[LCCF; e.g.,][]{welsh1999} in this work. We compute the cross-correlation coefficients with the following DCF and LCCF functions:
\begin{gather}
\textup{DCF}(\tau) = M^{-1}\,\left[\frac{\sum(a_{i} - \bar{a}_{0})(b_{j} - \bar{b}_{0})}{\sigma_{a\,0}\,\sigma_{b\,0}}\right], \label{eq:dcf1} \\
\textup{LCCF}(\tau) = M^{-1}\,\left[\frac{\sum(a_{i} - \bar{a}_{\tau})(b_{j} - \bar{b}_{\tau})}{\sigma_{a\,\tau}\,\sigma_{b\,\tau}}\right], \label{eq:lccf1}
\end{gather}
where $a_{i}$ \& $b_{j}$ being two time series, $\bar{a}_{0}$ \& $\bar{b}_{0}$ the mean values for the time series, $\sigma_{a\,0}$ \& $\sigma_{b\,0}$ the standard deviation values for the time series, and $M$ the number of ($i,j$) pairs that fall within
% a range of time delay ($\tau$)
\rv{the delay bin}
defined as $\tau\,-\,\Delta t/2\,\leq\,\Delta t_{ij}\,<\,\tau\,+\,\Delta t/2$;
% ($\Delta t$: \rv{a width of $\tau$}).
\rv{here $\tau$ and $\Delta t$ refer to time delay and a size of the delay bin, respectively.}
$\bar{a}_{\tau}$, $\bar{b}_{\tau}$, $\sigma_{a\,\tau}$, and $\sigma_{b\,\tau}$ are the same as above, but only for those $M$ overlapping samples.
We determine the uncertainties i.e., $\sigma_{\rm DCF}(\tau)$ and $\sigma_{\rm LCCF}(\tau)$ by measuring a statistical dispersion for each DCF($\tau$) and LCCF($\tau$), respectively \citep[see e.g.,][]{maxm2014b}. The main difference between the two approaches is that LCCF only considers data points overlapping within
% a delay width ($\Delta t$)
\rv{the delay bin}
between two time series. 

\begin{figure}[t]
\centering
\includegraphics[angle=0, width=\columnwidth, keepaspectratio]{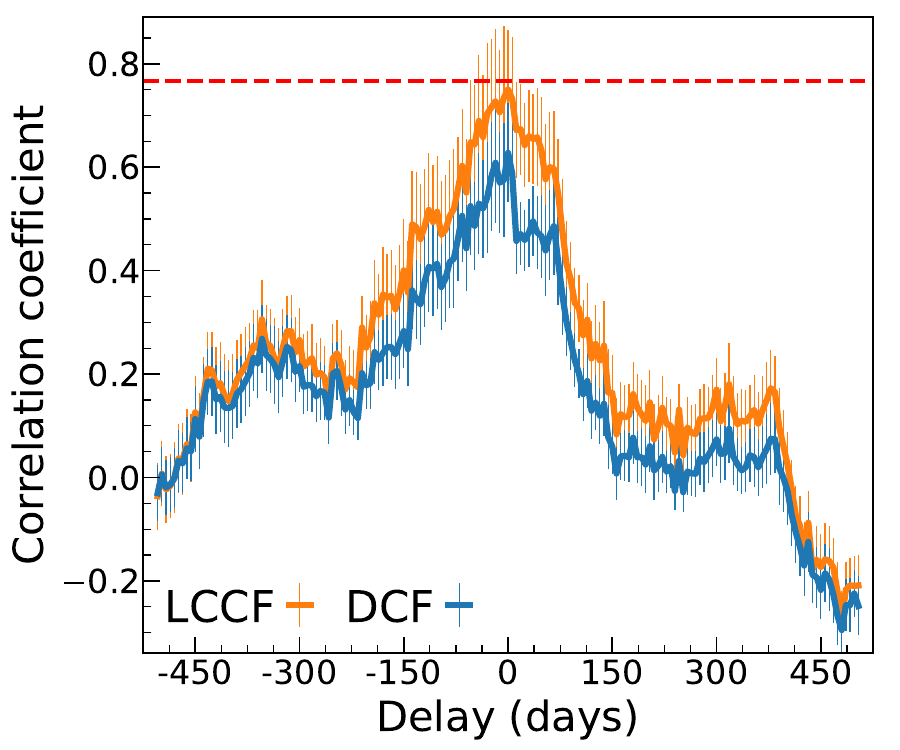}
\caption{
Comparison between LCCF and DCF. The $\gamma$-ray and ALMA band 3 light curves shown in Figure~\ref{fig:thelcs} were used in this test ($\Delta t$ = 6 days for both the correlation curves). The red dashed line indicates the average r$_{\rm P}$ value of Figure~\ref{fig:coreps}.
}
\label{fig:ccfcomp}
\end{figure}

\citet{maxm2014b} made a comparison between DCF and LCCF, and concluded that LCCF is better than DCF in terms of e.g., detection efficiency. For more details of the comparison, we refer to \citet{maxm2014b}.
We also tested these two functions with our datasets. Figure~\ref{fig:ccfcomp} shows the result. As evident from it, the LCCF curve returns higher coefficient values. Both the LCCF and DCF curves peaked at $\tau$ = 0. However, we find that the peak of the LCCF curve is much closer to the mean of the Pearson coefficients presented in Figure~\ref{fig:coreps}. Thus, we consider that LCCF is a more accurate way to search for time-correlation in this study.

We evaluated the resultant LCCF curves by using their confidence levels. The overall procedure is similar to the manner of \citet{kim2022}. We simulated 100,000 artificial $\gamma$-ray light curves and computed LCCF curves between those simulated $\gamma$-rays and the observed ALMA band 3 light curve. For every $\tau$, we draw a cumulative distribution function (CDF) of the correlation coefficients and find the 68\%, 95\%, and 99.9\% confidence levels. Based on these confidence levels, we determine the significance of the LCCF peaks.

\begin{figure}[t]
\centering
\includegraphics[angle=0, width=\columnwidth, keepaspectratio]{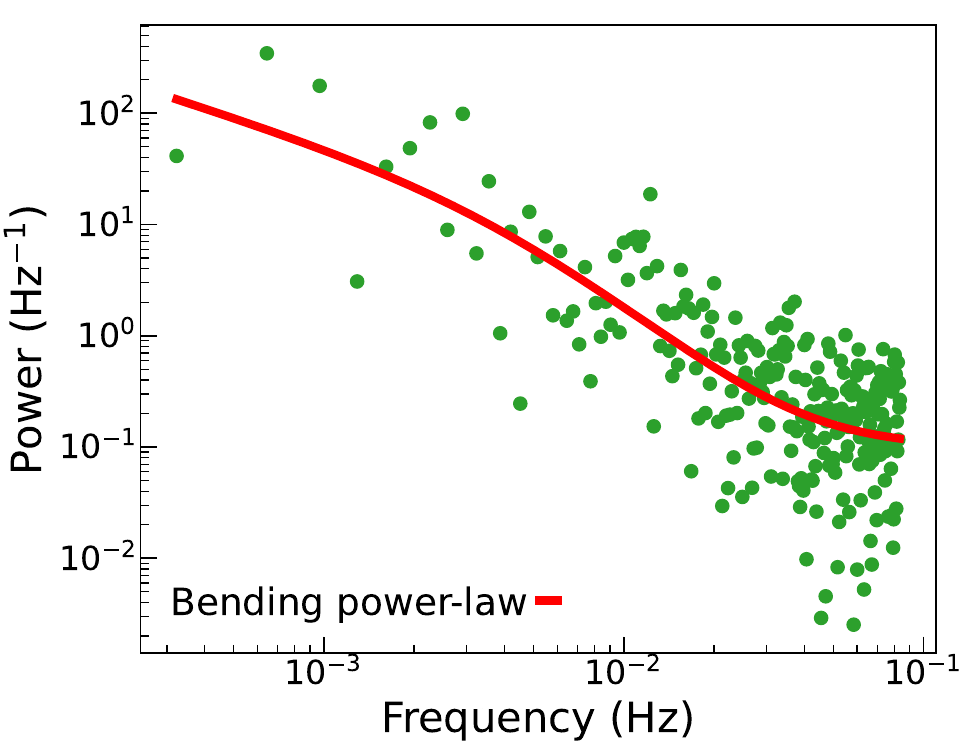} \
\includegraphics[angle=0, width=\columnwidth, keepaspectratio]{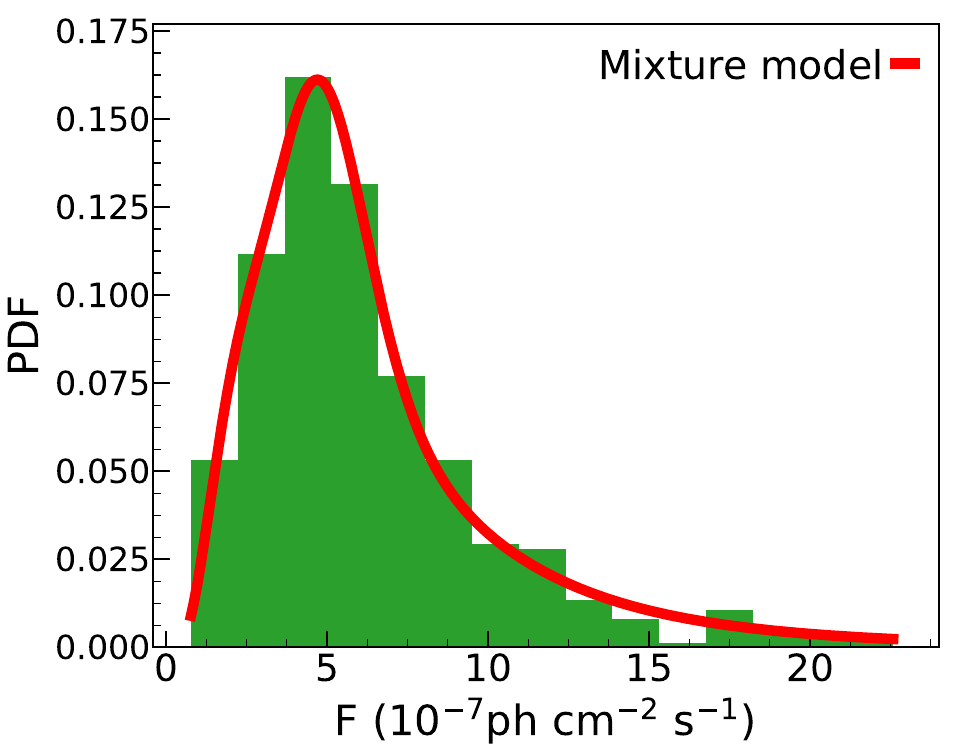}
\caption{
Best-fit PSD (\textsl{upper}) and PDF (\textsl{lower}) models (in red color) of the $\gamma$-rays.
}
\label{fig:gpsdpdf}
\end{figure}

% PSD PDF
The artificial $\gamma$-ray light curves were generated in the manner of \citet{emma2013}.
Most of the time, \object{PKS\,1424$-$418} was bright at $\gamma$-rays and there are a small number of bad/empty bins. Using linear interpolation, we complemented these bins and thus made the light curve evenly distributed. Then, we find a power spectral density (PSD) and a probability density function (PDF) of the $\gamma$-rays with this interpolated light curve. A bending power-law model and the mixture model (i.e., a combination of the gamma and log-normal distributions) introduced by \citet{emma2013}, were applied to a periodogram and histogram of the data, respectively. Figure~\ref{fig:gpsdpdf} shows the best-fit PSD and PDF models. Then, the artificial $\gamma$-ray light curves were generated with a Python implementation \texttt{DELCgen} \citep{conoly2015} that follows the algorithm of \citet{emma2013}, by using the estimated PSD \& PDF parameters
\footnote{PSD: $A=0.16\pm1.20$\,Hz$^{-1}$, $f_{\rm bend}=0.004\pm0.011$\,Hz, $\alpha_{\rm low}=0.84\pm1.04$, $\alpha_{\rm high}=2.22\pm0.99$, $C=0.10\pm0.04$\,Hz$^{-1}$, and PDF: $\kappa=17.94\pm9.42$, $\theta=0.30\pm0.15$, $\mu=1.70\pm0.03$, $\sigma=0.67\pm0.04$, $w_{\Gamma}=0.20\pm0.09$. For their mathematical forms, we refer to equations~2 \& 3 in \citet{emma2013}.}. 
Finally, we modified the sampling of all the artificial $\gamma$-ray light curves to make them the same as the observed $\gamma$-ray light curve in the time domain.

%\section{Tests for larger delay bins in LCCF}
\section{Tests on the LCCF curves}
\label{sec:ccf2}
\begin{figure*}[t]%[!hb]
%\begin{figure}[t]
\centering
\includegraphics[angle=0, width=60.5mm, keepaspectratio]{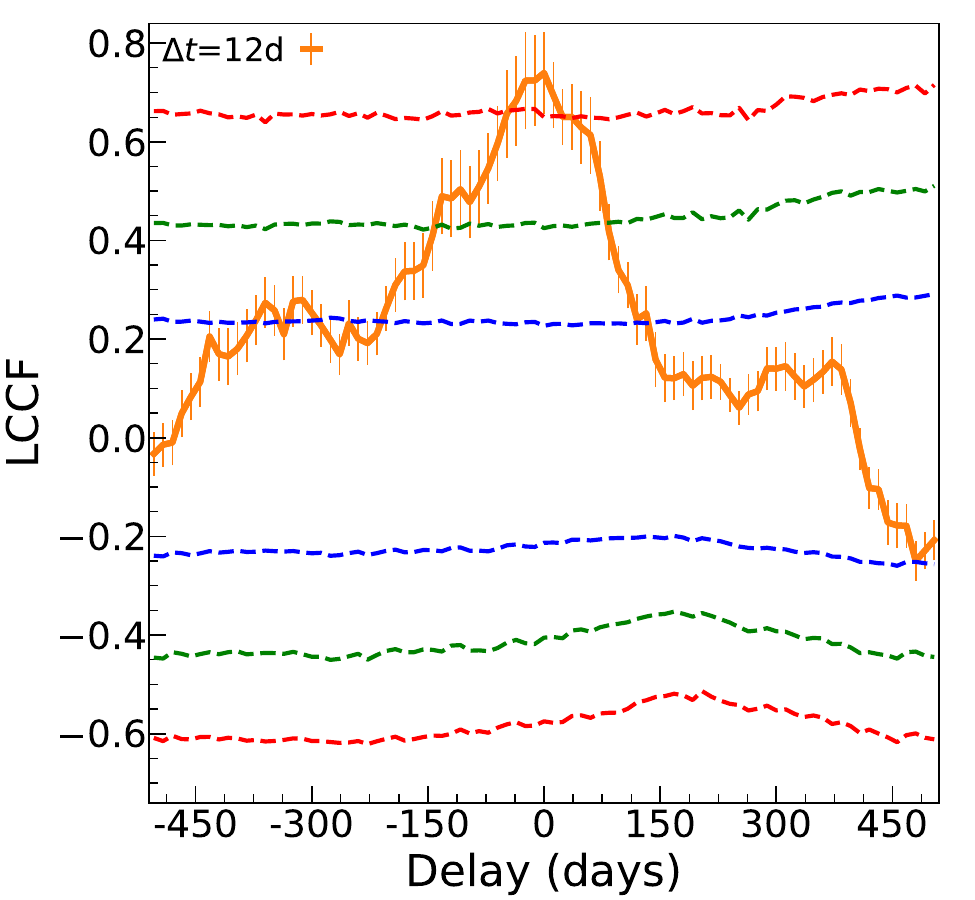} 
\includegraphics[angle=0, width=60.5mm, keepaspectratio]{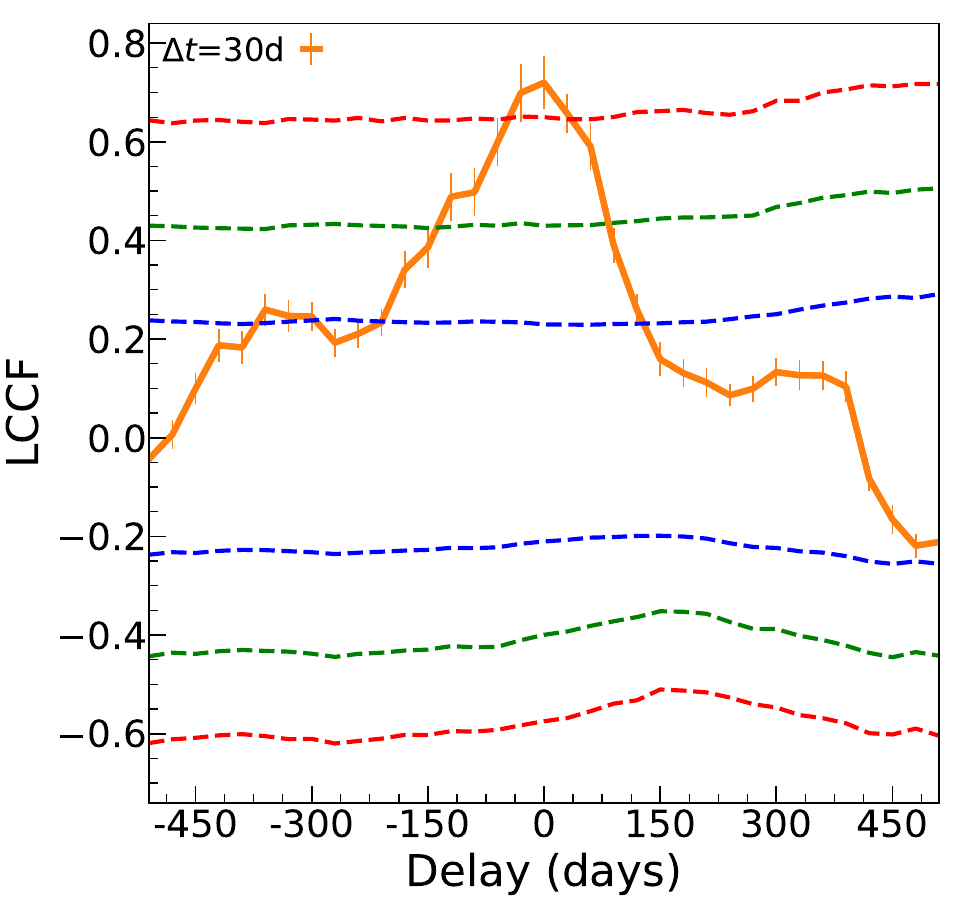} 
\includegraphics[angle=0, width=60.5mm, keepaspectratio]{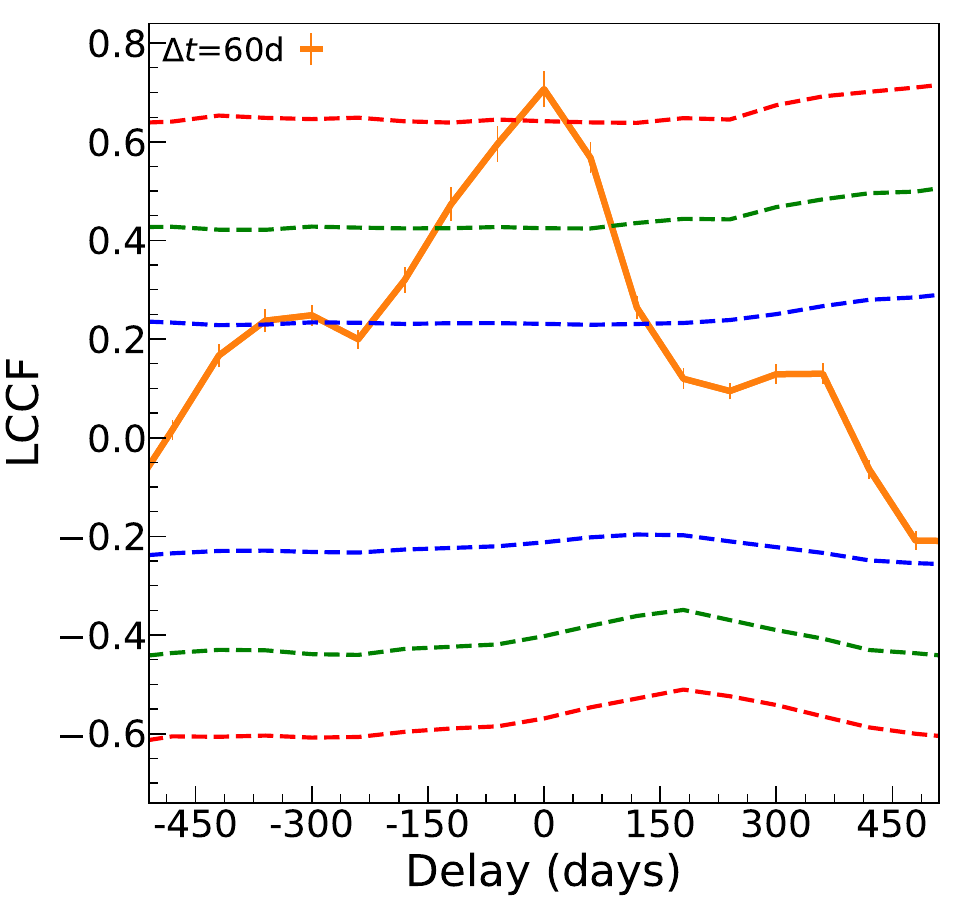}
\caption{
LCCF curves (in orange color) between the ALMA band 3 and $\gamma$-ray light curves of \object{PKS\,1424$-$418} over the whole 8.5\,yr. From left to right: $\Delta t$ = 12, 30, and 60\,days. The blue, green, and red dashed lines are the 68\%, 95\%, and 99.9\% confidence levels.
}
\label{fig:othlccf}
%\end{figure}
\end{figure*}

Figure~\ref{fig:othlccf} shows the LCCF curves between the $\gamma$-ray and ALMA band 3 light curves with larger $\Delta t$ values (i.e., 12, 30, and 60\,days). We find again a significant (i.e., $>$\,99.9\%), strong \rvc{mm}--$\gamma$-ray correlation at around $\tau$\,=\,0 (zero-lag). The overall shape of the LCCF curves are consistent with each other. However, those statistical fluctuations along the curves become smoother and weaker with increasing $\Delta t$. Some of the small humps (e.g., the one at around $\tau$\,$\sim$\,$-$120\,days) completely disappear with $\Delta t$\,=\,60\,days. They could be artifacts or short-term, local correlations. Thus, higher $\Delta t$ values could be helpful to identify a long-term correlation that remains throughout the whole period of the data.

\section{Gaussian fit to the LCCF curves}
\begin{figure}[hb]
\centering
\includegraphics[angle=0, width=82mm, height=67mm]{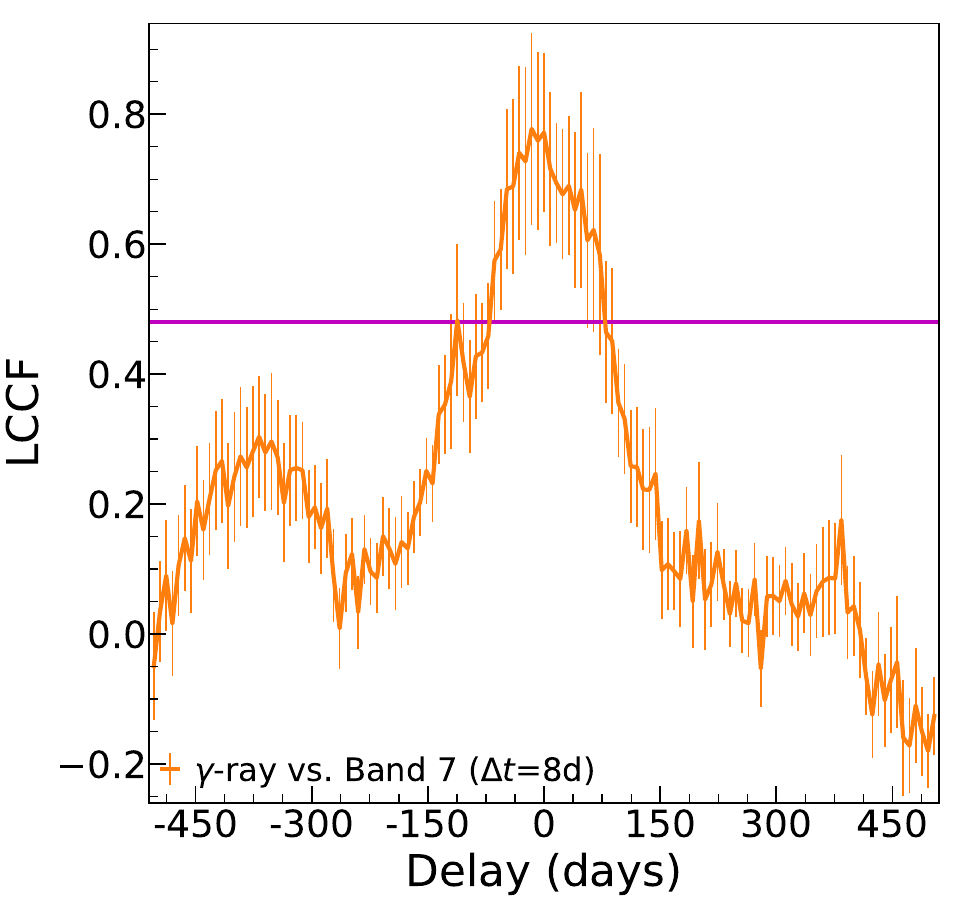}
\caption{
\rve{LCCF curve ($\Delta t$\,=\,8\,days) between the $\gamma$-ray and ALMA band 7 light curves, over the whole 8.5\,yr. The data points above LCCF\,$\sim$\,0.48 (i.e., the purple horizontal solid line) were used in the fit.
}
}
\label{fig:theb7curve}
\end{figure}

\begin{figure}[b]
\centering
\includegraphics[angle=0, width=82mm, height=67mm]{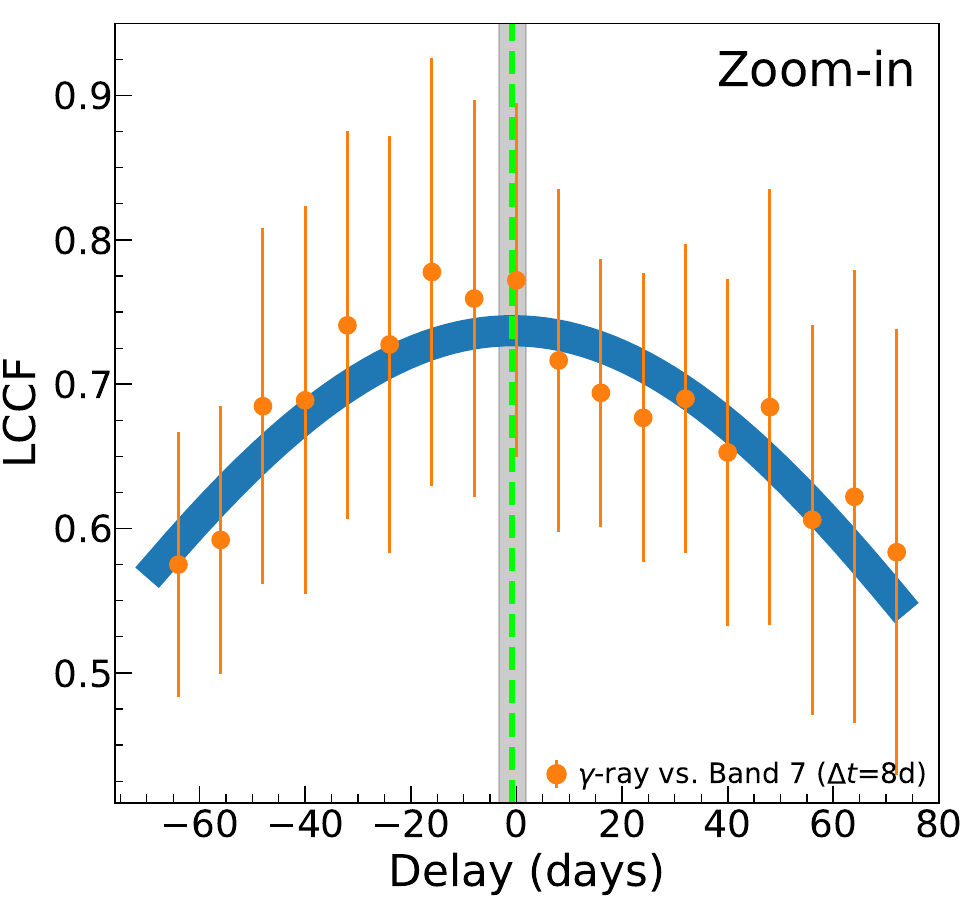} \
\includegraphics[angle=0, width=82mm, height=67mm]{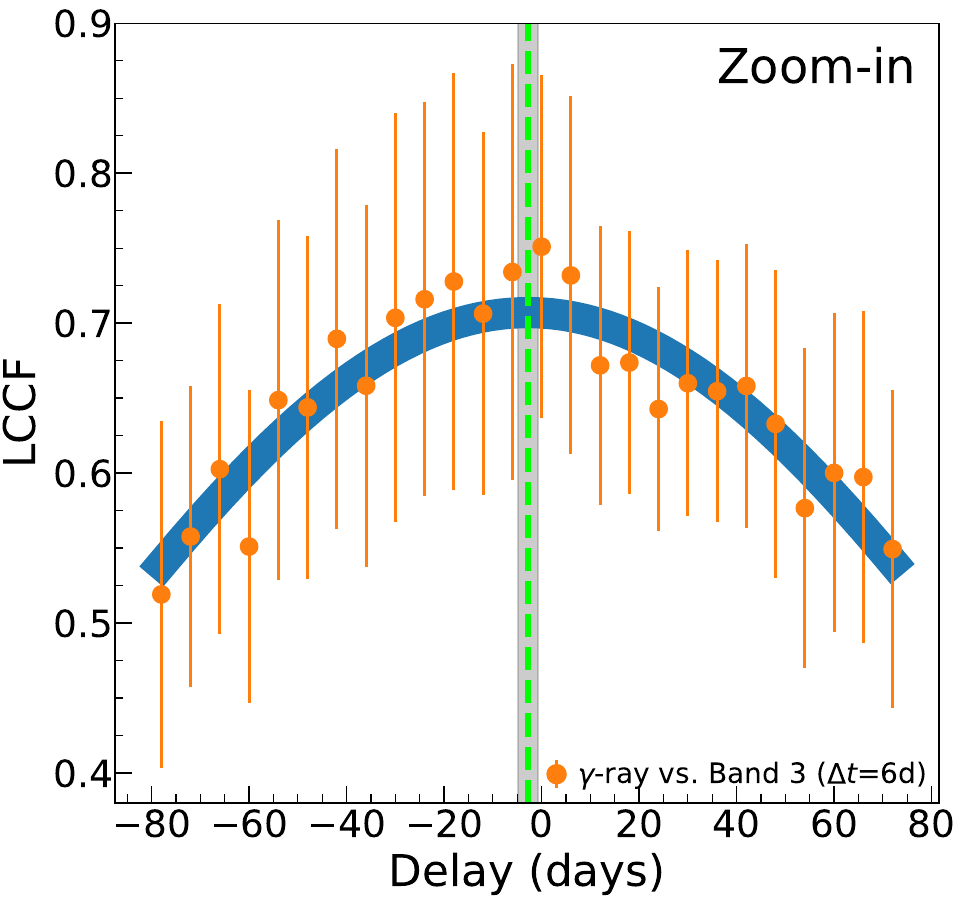}
\caption{
\rve{Top: Zoom-in-view of Figure~\ref{fig:theb7curve} shows the details of the fit results. The blue curve indicates the best-fit Gaussian model with a peak location of $-$0.8$\pm$2.6\,days (marked by the green dashed line with gray shaded area).
Bottom: Same as the above, but between the $\gamma$-ray and ALMA band 3 light curves. In Figure~\ref{fig:lccfmain}, the values above LCCF\,$\sim$\,0.52 were used in the fit. The estimated Gaussian peak is at $-$2.8$\pm$2.0\,days.
}
}
\label{fig:gausfits}
\end{figure}

% \texorpdfstring{$\gamma$}{g}
\label{sec:gausfit}
\rve{In Figure~\ref{fig:theb7curve}, we fit a single Gaussian to the LCCF curve between the LAT and ALMA band 7 light curves; we omitted the confidence levels for clarity, but the central hump is also above the 99.9\% level. $\Delta t$ was set to be 8\,days that is the median sampling of the band 7 light curve. To better fit the model to the data and focus on the significant central LCCF hump, we only used the LCCF points higher than the peak of a first strong, distinct side-lobe (i.e., LCCF\,=\,$\sim$0.48). The result can be found in the upper panel of Figure~\ref{fig:gausfits}.
We also estimate a peak position of the LCCF hump between the LAT and band 3 light curves (i.e., Figure~\ref{fig:lccfmain}), by using the same manner as Figure~\ref{fig:theb7curve}. We show the result in the lower panel of Figure~\ref{fig:gausfits}.}
Owing to the poor sampling of the band 6, we skip the band 6 data in the LCCF analysis; the statistical fluctuations and huge side-lobes occupy the LCCF curve between the LAT and band 6 data throughout the whole $\tau$ values.
\rva{We note that the Gaussian model well describes both the LCCF curves shown above, and the results are irrelevant to the bin size ($\Delta t$)}.

%Figure~\ref{fig:theb7curve}
%Figure~\ref{fig:gausfits}

%'vs. Band 7
%[ 0.73708549 -0.78206414 95.07347764]
%[0.01158232 2.60290117 6.62049157]
% theycut = 0.4830461466404257
%'vs. Band 3 
%[  0.7070671   -2.78918942 102.04172422]
%[0.00842028 2.02658551 5.18184326]

\end{appendix}
\end{document}